\documentclass[iop,apj,twocolappendix]{emulateapj}
\shorttitle{A {\it Chandra} survey of Galactic globular clusters}
\shortauthors{Cheng et al.}

\begin{document}
\title{A {\it Chandra} Survey of Milky Way Globular Clusters I: Emissivity and Abundance of Weak X-ray Sources}
\author{Zhongqun Cheng$^{1,2,3}$, Zhiyuan Li$^{2,3}$, Xiaojie Xu$^{2,3}$ and Xiangdong Li$^{2,3}$}
\affil{$^{1}$ Department of Astronomy, Xiamen University, Xiamen, Fujian 361005, China} 
\affil{$^{2}$ School of Astronomy and Space Science, Nanjing University, Nanjing 210023, China} 
\affil{$^{3}$ Key Laboratory of Modern Astronomy and Astrophysics (Nanjing University), Ministry of Education, Nanjing 210023, China}
\email{lizy@nju.edu.cn; lixd@nju.edu.cn}

\begin{abstract}
Based on archival {\it Chandra} data, we have carried out an X-ray survey of 69, or nearly half the known population of, Milky Way globular clusters (GCs), focusing on weak X-ray sources, mainly cataclysmic variables (CVs) and coronally active binaries (ABs).  
Using the cumulative X-ray luminosity per unit stellar mass (i.e., X-ray emissivity) as a proxy of the source abundance, we demonstrate a paucity (lower by $41\%\pm27\%$ on average) of weak X-ray sources in most GCs relative to the field, which is represented by the Solar neighborhood and Local Group dwarf elliptical galaxies.  
We also revisit the mutual correlations among the cumulative X-ray luminosity ($L_X$), cluster mass ($M$) and stellar encounter rate ($\Gamma$), finding $L_{X}\propto M^{0.74\pm 0.13}$, $L_{X}\propto\Gamma^{0.67\pm0.07}$ and $\Gamma \propto M^{1.28 \pm 0.17}$. 
The three quantities can further be expressed as $L_{X} \propto M^{0.64\pm0.12}\ \Gamma^{0.19 \pm0.07}$, which indicates that the dynamical formation of CVs and ABs through stellar encounters in GCs is less dominant than previously suggested, and that the primordial formation channel has a substantial contribution. 
Taking these aspects together, we suggest that a large fraction of primordial, soft binaries have been disrupted in binary-single or binary-binary stellar interactions before they can otherwise evolve into X-ray-emitting close binaries,
whereas the same interactions also have led to the formation of new close binaries.
No significant correlations between $L_{X}/L_{K}$ and cluster properties, including dynamical age, metallicity and structural parameters, are found.
\end{abstract}

\keywords{binaries: close -- globular clusters: general -- X-rays: binaries -- cataclysmic variables}

\section{Introduction} \label{sec:intro}
Globular clusters (GCs) are aged and self-gravitationally bound systems that evolve with a high stellar density.
Since the launch of the first X-ray satellite, {\it Uhuru}, it has been recognized that the abundance (i.e., number per unit stellar mass) of luminous X-ray binaries (with luminosity $L_{\rm X} \gtrsim 10^{36} {\rm \ erg\,s^{-1}}$) in GCs is $\sim$100 times higher than in the Galactic field \citep{Clark1975,Katz1975}. 
This over-abundance was attributed to the efficient formation of low-mass X-ray binaries (LMXBs) by stellar dynamical interactions in the dense core of GCs, where an isolated neutron star can be captured by a main sequence star through tidal force \citep{fabian1975}, by a giant through collision \citep{sutantyo1975}, or by a primordial binary through exchange with one of the constituent stars \citep{hills1976}. 
The fundamental parameter quantifying these dynamical interactions is the so-called stellar encounter rate, $\Gamma$, which is related to the stellar density ($\rho$) and velocity dispersion ($\sigma$), as $\Gamma \propto \int {\rho}^{2}/\sigma$, an integral over the cluster volume. 

With its superb angular resolution and sensitivity, the {\it Chandra X-ray Observatory} has resolved a large number of low-luminosity ($L_{X} \lesssim 10^{34} {\rm\ erg\ s^{-1}}$) source in GCs. 
When deep HST optical/ultraviolet images are available, the majority of these weak X-ray sources are found to be cataclysmic variables (CVs) and coronally active binaries (ABs), the rest being quiescent LMXBs (qLMXBs) and millisecond pulsars (MSPs) \citep[e.g.,][]{grindlay2001,pooley2002a,edmonds2003,heinke2003,heinke2005, haggard2009,maxwell2012}.  
These stellar systems either are experiencing the drastical binary evolution stage (i.e., qLMXBs, CVs, ABs) or are the immediate remnants of close binaries (i.e., MSPs), hence their formation could also have been affected by dynamical processes in GCs. 
Indeed, previous work revealed a correlation between the number of detected X-ray sources, in particular CVs, and the stellar encounter rate \citep{pooley2003,heinke2006,maxwell2012}, which was interpreted as a dominant fraction of these sources originating from dynamical processes \citep{pooley2006}.

However, it remains an open question of which dynamical process(es) is primarily responsible for the above correlation. 
In GCs, binaries tend to sink to the dense cluster core due to equipartition of kinetic energy, where dynamical processes including two-body and three-body encounters would take place with competing effects: binaries can be created in two-body interactions, but also can be destroyed or modified in three-body interactions \citep{hut1992a}. 
In particular, by binary-single interactions soft binaries (with bound energy $|E_{b}|$ less than the average stellar kinetic energy $E_{k}$) tend to be softer or even disrupted, whereas hard binaries (with $|E_{b}| > E_{k}$) become harder\footnote{Similar processes will take place in four-body (binary-binary) interactions, provided that the GC binary fraction is sufficiently high \citep[]{mikkola1983,hut1992a,hut1992b,bacon1996}.}\citep{heggie1975,hills1975,hut1993}.
While it is generally accepted that the total number of primordial binaries in GCs would decrease with time under dynamical interactions, it is far less clear whether the same processes would have a net effect of producing or destructing close binary systems such as CVs and ABs. 
The abundance of weak X-ray sources in GCs relative to the field offers a crucial diagnostics to this problem.

To date, few studies exist to quantify the relative abundance of weak X-ray sources in GCs. 
Based on {\it ROSAT} observations, \citet{verbunt2001} showed that most GCs have lower cumulative X-ray emissivities (i.e., $L_{X}$ per unit stellar mass) than that of the old open cluster M67. 
In a {\it Chandra} study of $\omega$ Centauri (NGC\,5139), \citet{haggard2009} found that the abundance of CVs in this cluster is at least 2-3 times lower than that of the field. 
On the other hand, using the K-band specific rate of classical novae detected in external galaxies as a proxy, \citet{Townsley2005} estimated that the CV abundance of an old stellar population
is compatible with the number of CVs detected in 47 Tuc.
More recently, Ge et al.~(2015) compared the cumulative X-ray emissivities of four Galactic GCs (including 47 Tuc and $\omega$ Cen) with the stellar X-ray emissivity averaged over several Local Group dwarf elliptical galaxies as well as the Solar neighborhood (Sazonov et al.~2006), but no firm conclusion could be drawn due to their limited GC sample.

In the present work, we study the largest sample of Galactic GCs observed by {\it Chandra} so far, to determine the abundance of X-ray sources and to examine its relation to various physical properties of the host cluster.
Unlike previous work (e.g., Pooley \& Hut 2006) that focused on the individually resolved sources, our approach, similar to Verbunt (2001) and Ge et al.~(2015), is to use the cumulative X-ray emissivity\footnote{The contribution of luminous LMXBs, if present, are excluded. See Section 2.1 for details.} as a proxy of the source abundance.
The source-counting method is inevitably subject to the strongly varied detection sensitivity among different GCs, resulting in a limited sample size. This method is also subject to contamination of foreground/background interlopers, which were often not properly accounted for. 
In contrast, our approach of measuring the total X-ray emissivity is generally insensitive to the exposure time or distance of a given GC, and thus in principle can be applied for a large GC sample in a highly uniform fashion. 
Moreover, the derived GC X-ray emissivities can be directly compared to the cumulative stellar X-ray emissivity of other galactic environments, e.g., the Galactic field and dwarf elliptical galaxies, which are crucial to evaluating the relative source abundance in GCs.

The limitation of our approach lies in that we do not distinguish the various X-ray populations (except that luminous LMXBs are precluded). However, it has been demonstrated that CVs and ABs together dominate the X-ray emission from GCs \citep{pooley2006,heinke2011}.  
For example, \citet{heinke2005} have detected $\sim$300 X-ray sources in NGC\,104 using deep {\it Chandra} observations. They estimated that roughly 70 are background sources, 5 are qLMXB candidates, 25 are MSP candidates, and the remaining majority ($\sim$200) are CVs and ABs. 
As we will show below, the average GC X-ray emissivity clearly indicates a paucity of X-ray sources. Any minor contribution by qLMXBs and MSPs to the measured emissivity only strengths our conclusion. 
Moreover, in practice there is no clear cut between CVs and ABs in their X-ray appearance, perhaps except that CVs are on average harder and more luminous.         
CVs and ABs are also closely related to each other from the veiwpoint of binary evolution.

The remainder of this paper is organized as follows. Section 2 describes data reduction and analysis that lead to a uniform measurement of the X-ray luminosity and stellar mass of individual GCs.  
Section 3 explores correlations between the X-ray luminosity or X-ray emissivity and various physical properties of the GCs. Discussion and summary of our results are given in Section 4 and Section 5, respectively.    
Throughout this work we quote 1\,$\sigma$ errors unless otherwise stated.

\section{Data Preparation and Analysis}    

\subsection{X-ray data and sample selection}

To date, 157 Galactic GCs have been discovered and tabulated in the catalogue of \citet{harris1996}. We searched the {\it Chandra} archive for all GCs with ACIS-I or ACIS-S observations taken by May 2014.
Some GCs are known to harbor luminous LMXBs that easily dominate the total X-ray emission from the host cluster. We consulted the LMXB catalogue of \citet{liu2007} and visually inspected the {\it Chandra} images for the presence of luminous LMXBs. 
We found that in most such cases even just the PSF-scattered halo of the LMXB would severely affect our analysis, and hence we decided to remove all GCs hosting luminous LMXBs from our sample.
Our final sample thus consists of 69 GCs (Table~\ref{tab:GC}), nearly half of the known Galactic GC population. Among them, 21 are dynamically old and have been designatated ``core-collapsed" GCs in the catalogue of \citet{harris1996}. We refer to the rest (48) as dynamically normal GCs.
Notably, the size of our sample is $\sim$6 and $\sim$3 times that investigated by \citet{pooley2003} and \citet{pooley2006}, respectively\footnote{The preliminary analysis of source abundance in \citet{pooley2010} considered 63 GCs, but a large fraction of this sample was necessarily binned by the encounter rate to ensure a significant number of detected sources in each bin.}. 

We used CIAO v4.5 and the corresponding calibration files to reprocess the data, following the standard procedure\footnote{http://cxc.harvard.edu/ciao}. Because a substantial fraction of the total X-ray flux of a given GC may come from its unresolved emission, care was taken to filter background flares using {\it lc\_clean}. 
A log of the $\it Chandra$ observations analyzed in this work is given in Table~\ref{tab:log}. 

\subsection{X-ray flux measurement} \label{subsec:xflux}

We took the following steps to measure the net flux for each GC. First, we defined the cluster and background regions in a uniform fashion. Due to the mass segregation effect, a large fraction of the X-ray sources are likely to concentrate within $2-3\,r_c$, where $r_c$ is the core radius (Harris, 2010 edition). 
We adopted the half-light circle, of radius $r_h$, as the cluster region for all sample GCs.
This is a reasonable choice for the dynamically normal GCs, in which $r_h$ is typically a few times of $r_c$. For core-collapsed GCs, the ratio of $r_h/r_c$ are much larger (Table~\ref{tab:GC}), thus contamination from foreground/background sources can be substantial. Nevertheless, any contribution from the interlopers would be statistically subtracted (see below). 
Our choice of the cluster region also facilitates direct comparison with previous work \citep{pooley2003,pooley2006}.

\begin{figure*}[hbtp]
\centering
\includegraphics[width=1.0\textwidth]{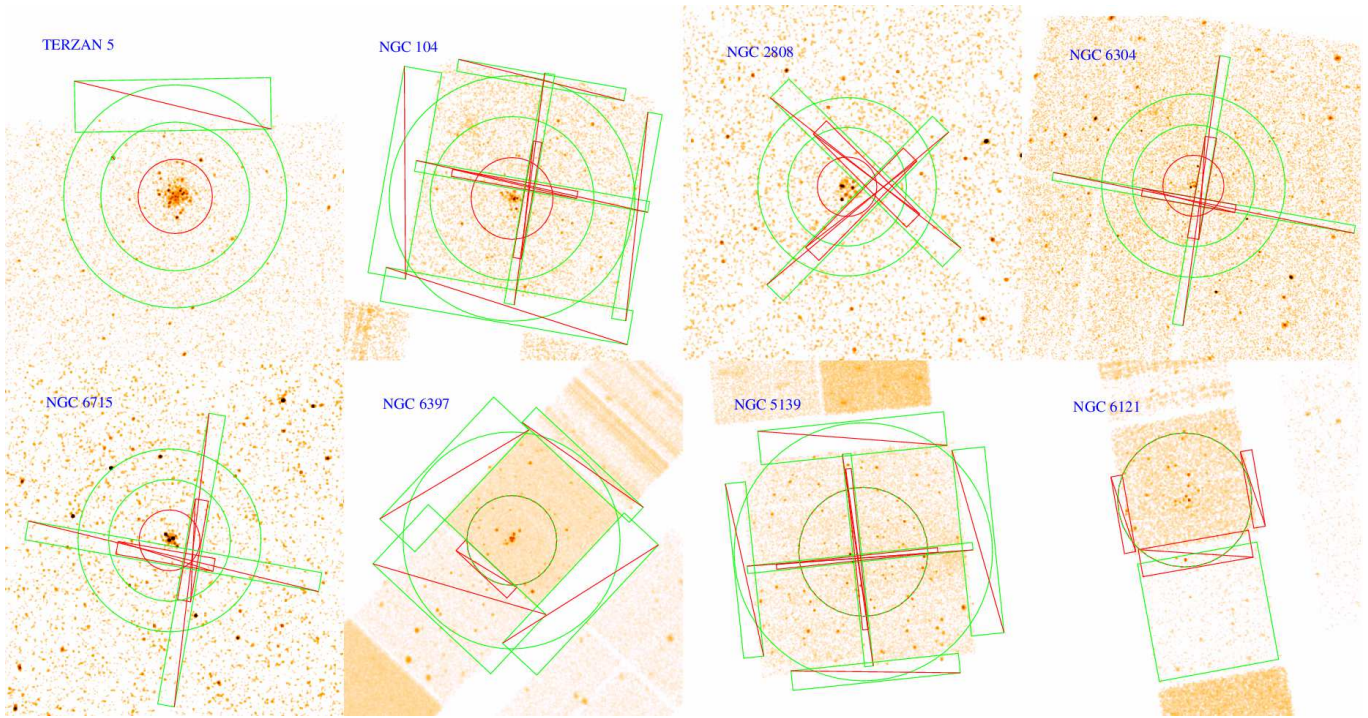}
\linespread{0.7}
\caption{Examples of GCs with large source extraction regions. We defined for each GCs the source region ($S_{h}$) as a (red) circle with radii of $r_h$, while defined the background region as a (green) annulus with inner-to-outer radii of $2r_h-3r_h$. Regions near CCD edges or gaps (rectangles with red diagonal lines) are masked to minimize uncertainty in the detector response.  
For GCs with a large $r_h$, the background annulus was set as $r_h-2r_h$ and from the same CCDs as for the source region; the only exception is NGC\,6121, where the background region was chosen from the S2 CCD (green rectangle in the lower right panel). We adopted a `double subtraction' procedure to correct for the vignetting effect for these GCs (see text for details). \label{fig:regions}}
\end{figure*}

Since the {\it Chandra} field-of-view (FoV) is not always large enough to cover the half-light circle (Figure~\ref{fig:regions}), we defined for each observation a correction parameter $k_{1}=S_{h}/{S_{\rm FoV}}$, where $S_{h}=\pi r_{h}^{2}$ and $S_{\rm FoV} \leq S_{h}$ is the actual area allowed by the FoV. 
In this regard, the total flux is calculated as $F_{\rm X}=k_{1}\times F_{\rm X, FoV}$, where $F_{\rm X, FoV}$ is the flux within $S_{\rm FoV}$. 

For the background region, we adopted an annulus with inner-to-outer radii of $2r_h-3r_h$ when the FoV was large enough, but for those GCs with a large $r_h$, the annulus was set as $r_h-2r_h$. 
We required that the background region fall within the same CCDs (AICS-I or ACIS-S3) as of the cluster region, but for NGC 6121, the background was chosen from ACIS-S2. We also avoided CCD edges and gaps where detector response may have a substantial uncertainty. We defined for each observation a second parameter $k_{2}=S_{\rm FoV}/{B_{\rm FoV}}$, where $B_{\rm FoV}$ is the actual background area allowed by the FoV (Figure~\ref{fig:regions}). In most cases, $k_{2}$ was less than $0.5$ (Table~\ref{tab:log}).

Next, we calculated for each observation the cluster net counts as $N_S-N_B'=N_S-k_2N_B$, where $N_S$ and $N_B$ are the 0.5--8 keV counts within $S_{\rm FoV}$ and $B_{\rm FoV}$, respectively.
The signal-to-noise ratio, $S/N=(N_S-N_B')/\sqrt{N_S+N_B'}$, was derived for each observation and listed in Table~\ref{tab:LX}.
Among the 69 GCs, 51 have $S/N \geq 3$ and are considered as solid detections.

For most of our sample GCs, the total X-ray luminosity is not expected to be dominated by any single source. However, a source caught in a rare outburst may affect the total X-ray luminosity substantially. 
Therefore, for each observation we ran the CIAO {\it wavdetect} script to detect sources. 
A source located in $S_{\rm FoV}$ was referred to as an outbursting source if its 0.5-8 keV unabsorbed luminosity, derived with a power-law spectral model, is greater than $3\times10^{33}{\rm~erg~s^{-1}}$. 
Such bright sources, found in only 3 GCs (Table~\ref{tab:burst}), in which they each contribute more than $70\%$ of the total net count rate, were subsequently removed for flux calculation. 

Spectra were then extracted from the cluster and background regions for each GC. The background-subtracted cluster spectra were grouped to have at least 20 counts and a minimum signal-to-noise ratio of 3 per bin. 
We performed spectral analysis with Xspec v12.8.0. The models adopted to fit the spectra were either an absorbed power-law (phabs*powerlaw), an absorbed bremsstrahlung (pabs*brem), or a combination of the two (phabs*(powerlaw+brem)). 
We calculated the absorption column density ($N_{\rm H}$) of each GC from their color excess $E(B-V)$ and fixed this parameter in the fit. 
If a GC had more than one observations, a joint fit was performed to minimize the statistical uncertainty, allowing the normalization to vary but having the other parameters linked among the observations.
The 0.5--8 keV unabsorbed flux was derived from the best-fit model. 
For the 18 GCs with $S/N < 3$, we derived for them an upper limit in the net count rate (at $95\%$ confidence) using the CIAO tool {\it aprates}\footnote{{By assuming an non-informative prior distributions for the background and source intensity, \it aprates} uses Bayesian statistics to compute the background-marginalized, posterior probability distribution for source intensity, which can be used to determine the intensity value and confidence bounds or intensity upper limit.}, and converted it into an unabsorbed flux assuming an intrinsic power-law model with the photon-index fixed at $2.0$. 
We then calculated the 0.5--8 keV luminosity or upper limit by adopting the cluster distance from \citet{harris1996}. Our spectral analysis results are summarized in Table~\ref{tab:LX}.

Direct subtraction of the local background is expected to be a sufficient treatment for most GCs. For 13 GCs with a relatively large angular extent ($r_h \gtrsim 2'$; marked by ``*" in Table~\ref{tab:LX}), however, 
the effect of vignetting may bias low the background level, which is necessarily estimated at large off-axis angles.   
For such GCs, we corrected for vignetting following the `double-subtraction' procedure (e.g., Li et al.~2011): a first subtraction of the non-vignetted instrumental background followed by a second subtraction of the vignetted cosmic background. 
Briefly, we first generated the instrumental background spectra for both the cluster and background regions, using the {\it Chandra} ``stowed" background files. 
We then characterized the local cosmic background spectrum (i.e., instrumental background-subtracted) with a phenomenological model, being either an absorbed power-law or a power-law plus APEC thermal plasma with absorption. 
The hence derived cosmic background was added as a fixed component to the total spectral model, after scaling with the sky area. 

Lastly, we examined the effect of source variability on the derived total X-ray luminosity, focusing on three GCs (Terzan 5, NGC 6626 and NGC 6397) with multiple observations. The spectra extracted from individual observations were fitted independently with an absorbed power-law model. 
The results, listed in Table~\ref{tab:var}, indicate that in all three cases the derived X-ray luminosity varies by less than $30 \%$ over a timespan of years. This mild variability should have little effect on our statistical analysis and conclusions below.

\subsection{2MASS Data} \label{subsec:2MASS} 

To obtain an accurate measurement of the X-ray emissivity, a well determined cluster stellar mass is required. 
We adopted the K-band image of the Two Micron All Sky Survey (2MASS; \citealt{jarrett2000}), which is expected to be a better proxy of old stellar populations in GCs than the optical bands such as those quoted in \citet{harris1996}. 
We downloaded the archival 2MASS images of the individual GCs and derived their K-band luminosity, $L_{K}$, as follows. The cluster and background regions for photometry were the same as those used in the X-ray analysis (Section~\ref{subsec:xflux}). 
If a single 2MASS image did not cover the defined regions, we calibrated several adjacent images and merged them into a mosaic image with astrometry correction. 
We then obtained the K-band apparent magnitude of the cluster region after subtracting the background and correcting for extinction with $E(B-V)$. 
Finally, we calculated $L_{K}$ according to the apparent magnitude and cluster distance, which is expressed in units of solar K-band luminosity $L_{K,\odot}=5.67\times10^{31}$ $\rm erg\,s^{-1}$ (for solar K-band absolute magnitude $M_{K,\odot}=+3.31$). 
The results are listed in Table~\ref{tab:LX}.

\begin{figure}[htbp]
\centering
\includegraphics[width=.5\textwidth]{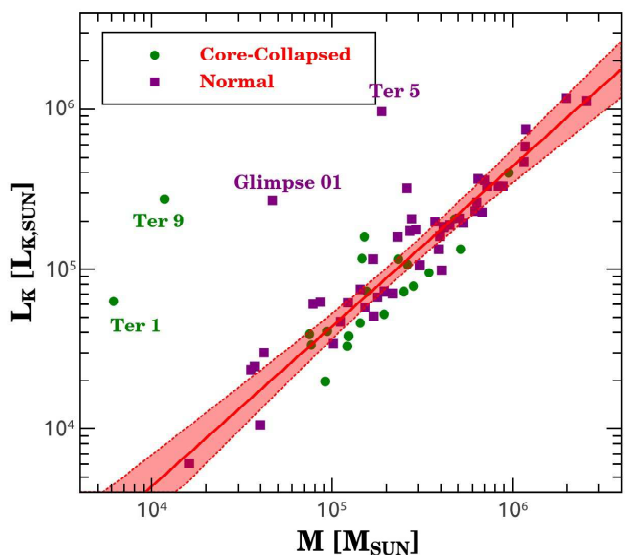}
\caption{GC K-band luminosity versus cluster mass $M$, which is derived from V-band magnitude following the empirical relation of \citet{majun2015}. 
The olive dots and purple squares donate the core-collapsed and dynamically normal GCs, respectively. 
The red line is the best-fit function, with shaded area representing the $95\%$ confidence range. The four outliers are marked. \label{fig:K-M}}
\end{figure} 

To establish the use of $L_{K}$ as a proxy of the cluster mass $M$, we first calculated $M$ using the V-band magnitude listed in \citet{harris1996}, following the empirical relation by \citet{majun2015}, $M/{\rm M_{\odot}}=10^{-0.4(M_{V}-5.764 \pm 0.063)}$, which is based on multi-band photometry of 297 GCs in M\,31 and stellar population synthesis models. 
Figure~\ref{fig:K-M} shows that $L_{K}$ is tightly correlated with $M$. Most GCs follow a fitted power-law function $L_{K}/L_{K,\odot}=10^{-0.38\pm 0.23} \times {(M/{\rm M_{\odot}})}^{1.00 \pm 0.04}$, which suggests a quasi-linear relation between $L_{K}$ and $M$. 

In the above fit, we have excluded four significant outliers (Terzan1, Terzan 5, Terzan 9 and Glimpse 01). We note that all these four GCs suffer from strong foreground extinction, thus their V-band fluxes may have been underestimated. 
For example, the V-band magnitude-based masses of Terzan 5 and Glimpse 01 are $1.88 \times 10^{5} {\rm M_{\odot}}$ and $4.67 \times 10^{4} {\rm M_{\odot}}$, respectively, whereas their K-band luminosities, $9.67 \times 10^{5} L_{K,\odot}$ and $2.41 \times 10^{5} L_{K,\odot}$, predict 
$\sim$12 times larger masses according to the above $L_{K}-M$ relation. 
Hereafter, we adopt a cluster mass of $2\times10^{6} {\rm~M_{\odot}}$ for Terzan 5 and $3 \times 10^{5} {\rm~M_{\odot}}$ for Glimpse 01, which were obtained by \citet{lanzoni2010} and \citet{pooley2007}, respectively, and are consistent with our $L_{K}-M$ relation. 
Because there is no reliable source of cluster mass for Terzan 1 and Terzan 9, we do not include these two core-collapsed clusters in the following statistical analysis. 

\section{Statistical relations} \label{sec:stat}

As discussed in Section~\ref{sec:intro}, in the absence of luminous LMXBs, the bulk of X-ray emission from GCs arises from a collection of CVs, ABs, qLMXBs and MSPs, hence the cumulative X-ray luminosity ($L_{X}$) should be roughly scaled with the total number of such sources, and more fundamentally, scaled with the cluster mass ($M$). 
If dynamical interactions are prone to create X-ray sources, a correlation between $L_{X}$ and the stellar encounter rate ($\Gamma$) is also expected. 
Other physical properties, such as metallicity, stellar density, age and structural parameters, of the GCs may also affect the formation and evolution of the X-ray populations. 
In this section, we examine the dependence of the cumulative X-ray emissivity, and hence the source abundance, on the various cluster properties. 
We define the cumulative X-ray emissivity as $\epsilon_{X} = 2L_{X}/M$, where the factor of 2 accounts for the fact that $L_{X}$ as given in Table~\ref{tab:LX} has been measured within the half-light circle.
The quasi-linear $L_{K}-M$ relation derived in Section~\ref{subsec:2MASS} also allows us to adopt the quantity $L_{X}/L_{K}$ as a proxy of the X-ray emissivity, which has the virtue of being distance-independent.
The GC parameters are taken from Harris (2010 edition); errors, when available and relevant, are also quoted. 

\subsection{Correlations with cluster mass}

\begin{figure}[htbp]
\centering
\includegraphics[width=.5\textwidth]{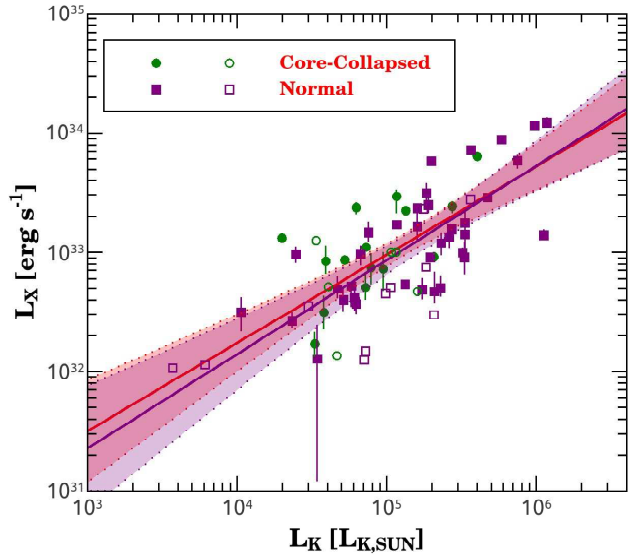}
\linespread{0.7}
\caption{GC X-ray luminosity as a function of the K-band luminosity. The olive circles and purple squares donate the core-collapsed and dynamically normal GCs. Filled and open symbols represent the luminosity and upper limit, respectively. The solid lines mark the best-fitting functions, purple for the dynamically normal GCs and red for the full sample, respectively. The shaded area represents the $95\%$ confidence range. \label{fig:X-K}}
\end{figure}

In Figure~\ref{fig:X-K}, we plot $L_{X}$ versus $L_{K}$ for all GCs. 
A correlation between $L_{X}$ and $L_{K}$ is evident, for which we find the Spearman's rank correlation coefficient $r=0.694 \pm 0.078$, with the p-value of $7.4 \times 10^{-11}$ for random correlation. 
We fit a power-law function to the correlation and obtain $L_{X} \propto L_{K}^{0.74 \pm 0.13}$, which is marked by the red solid line in Figure~\ref{fig:X-K}. 
Here and below we exclude the GCs with $S/N < 3$ (highlighted with open symbols in the figures) from the fit to the correlations.
There is no significant difference between the dynamically normal and core-collapsed GCs. A power-law fitting function for the dynamically normal GCs gives $L_{X} \propto L_{K}^{0.79 \pm 0.12}$ (purple line in Figure~\ref{fig:X-K}).
Notably, the fitted indice imply a sub-linear correlation.
      
\begin{figure}[htbp]
\centering
\includegraphics[width=.5\textwidth]{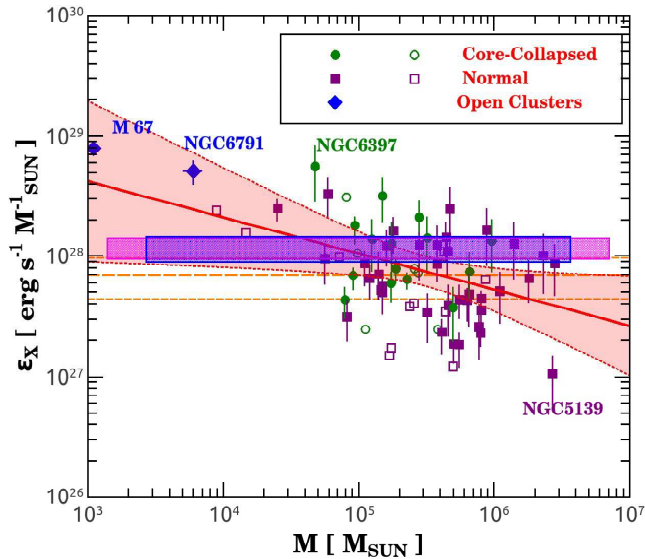}
\linespread{0.7}
\caption{X-ray emissivity as a function of the cluster mass, which is derived from the best-fit $L_{K}-M$ relation shown in Figure~\ref{fig:K-M}. 
The olive circles and purple squares denote the core-collapsed and dynamically normal GCs. Filled and open symbols represent the measured X-ray emissivity and upper limit, respectively. The red solid line delineates the best-fitting function, with shaded area represents the $95\%$ confidence range, while the dashed line represents the mean X-ray emissivity of the full sample (dispersions marked by the pair of dashed lines). 
For comparison, the cumulative X-ray emissivities of CVs and ABs in the Solar vicinity \citep{Sazonov2006,Revnivtsev2007} and of the Local Group dwarf elliptical galaxies \citep{ge2015} are marked by the blue and magenta strips, respectively.
Two Galactic old open clusters from \cite{vandenberg2013} are marked by blue diamonds, which were not included in the fit. See text for details. \label{fig:em}}
\end{figure}

In Figure~\ref{fig:em}, we plot the GC X-ray emissivity $\epsilon_{X}$ versus $M$. Here the latter has been derived from $M/{\rm M_{\odot}}=10^{0.38} \times (L_{K}/L_{K,\odot})$ rather than the V-band magnitude-based masses (Section 2.2). 
It can be seen that $\epsilon_{X}$ has a substantial scatter ranging from $10^{27}$ to a few times $10^{28}\rm\ erg\ s^{-1}{\rm M^{-1}_{\odot}}$. 
Nevertheless, a marginally significant negative correlation between $\epsilon_{X}$ and $M$ for the full sample is suggested by the Spearman's rank correlation coefficient $r=-0.323 \pm 0.077$, with $p=0.007$ for random correlation. 
The best fitting power-law function is $\epsilon_{X} \propto M^{-0.30 \pm 0.11}$ (red solid line in Figure~\ref{fig:em}). 
However, if NGC\,6397 and NGC\,5139 (labelled in Figure~\ref{fig:em}) are not included in the fit, the $\sim$ $3\,\sigma$ anti-correlation becomes $\sim$ $2\,\sigma$, with $\epsilon_{X} \propto M^{-0.19 \pm 0.10}$.
We note that this anti-correlation is consistent with the sub-linear correlations found in Figure~\ref{fig:X-K}.

We also measure the average X-ray emissivity of all 67 GCs to be $(6.9\pm2.6)\times10^{27}\rm\ erg\,s^{-1}M^{-1}_{\odot}$, as marked in Figure~\ref{fig:em} by the dashed horizantal lines. If only GCs with $S/N\geq 3$ were considered, the average emissivity is $(7.4\pm2.8)\times10^{27}\rm\ erg\,s^{-1}M^{-1}_{\odot}$.
If the three outburst sources were taken back into account, the X-ray emissivity of their host GCs would increase substantially (Table~\ref{tab:burst}), but they have little effect in the average GC emissivity, which becomes $(7.6\pm2.7)\times10^{27}\rm\ erg\,s^{-1}M^{-1}_{\odot}$. 
The average X-ray emissivities of the core-collapsed and dynamically normal GCs are marginally consistent with each other ($9.4\pm3.4$ versus $6.2\pm2.3\times10^{27}\rm\ erg\,s^{-1}M^{-1}_{\odot}$).

For comparison, also shown as a blue strip in Figure~\ref{fig:em} is the cumulative X-ray emissivity of CVs and ABs detected in the Solar neighborhood, with a value of $\epsilon_{X} = (11.7 \pm 3.1) \times 10^{27}\rm\ erg\,s^{-1}M^{-1}_{\odot}$, which was based on the X-ray luminosity function of CVs and ABs with $L_{X}$ ranging from $10^{27}$ to $10^{34}\rm\ erg\,s^{-1}$\citep{Sazonov2006}\footnote{To derive the 0.5--8 keV emissivity, we have converted Sazonov et al.'s 2--10 keV emissivity, $(3.1 \pm 0.8) \times 10^{27}\rm\ erg\,s^{-1}M^{-1}_{\odot}$), into the 2--8 keV band, by assuming a power-law spectrum with a photon-index of 2.1, suitable for the Galactic Ridge X-ray emission, and added the 0.5--2 keV emissivity, $(9 \pm 3) \times 10^{27}\rm\ erg\,s^{-1}M^{-1}_{\odot}$), as given in \citet{Revnivtsev2007}.}.
Also compared in Figure~\ref{fig:em} (magenta strip) is the average stellar X-ray emissivity of four gas-poor dwarf elliptical galaxies (M\,32, NGC\,147, NGC\,185 and NGC\,205), with a value of $\epsilon_{X}= (11.9 \pm 2.7) \times 10^{27}\rm\ erg\,s^{-1}M^{-1}_{\odot}$ derived from \cite{ge2015}. 
In their work, discrete sources with X-ray luminosities greater than $\sim$ $10^{34}\rm\ erg\,s^{-1}$ have been removed to ensure that the unresolved X-ray emission from these galaxies is dominated by CVs and ABs. 
We note that the Solar neighborhood and the dwarf elliptical galaxies exhibit highly similar stellar X-ray emissivities, indicating a quasi-universal emissivity in normal galactic environments, i.e., where stellar dynamical effects are not important \citep{ge2015}. 
On the other hand, as clearly shown in Figure~\ref{fig:em}, most GCs have a cumulative X-ray emissivity lower than the field level represented by the Solar neighborhood and the dwarf ellipticals. {\it This strongly suggests a dearth rather than over-abundance of weak X-ray sources in GCs relative to the field \footnote{The power-law luminosity function $dN\propto L_{X}^{-\gamma}d \rm ln\ L_{X}$ of GC X-ray sources has a typical slope of $\gamma <1$ \citep{pooley2002b}, while the slope of X-ray sources in the Solar neighborhood is $\gamma \approx 1.22$ in the luminosity range $10^{30}-10^{34}\rm\ erg\,s^{-1}$ \citep{Sazonov2006}. Therefore, this arguement is insensitive to the number of X-ray sources at the faint-end.}}. 

For completeness, we place in Figure~\ref{fig:em} two old open clusters, M\,67 and NGC\,6981. Their cumulative X-ray emissivities, derived from \cite{vandenberg2013}, are significantly higher than that of most GCs and the field. 
Such a trend was previously noted by Verbunt (2001) based on ROSAT observations, and by Ge et al.~(2015) based on only four GCs observed by {\it Chandra}.

\subsection{Correlations with stellar encounter rate}

Traditionally, the stellar encounter rate, $\Gamma \propto \int {\rho}^{2}/\sigma$, has been evaluated as $\Gamma \propto \rho_{c}^{2} r_{c}^{3}/\sigma_{c}$ or $\Gamma \propto \rho_{c}^{1.5} r_{c}^{2}$ \citep{verbunt1987,verbunt2003}, with the assumption that the spatial distribution of stars in GCs can be characterized by the King model \citep{king1962,1966}.
Here $r_{c}$ is the core radius and $\rho_{c}$ the average stellar density within $r_{c}$. The central velocity dispersion follows $\sigma_{c} \propto \rho_{c}^{0.5} r_{c}$ for a virial system.
 
\begin{figure*}[htbp]
\centering
\includegraphics[width=0.5\linewidth]{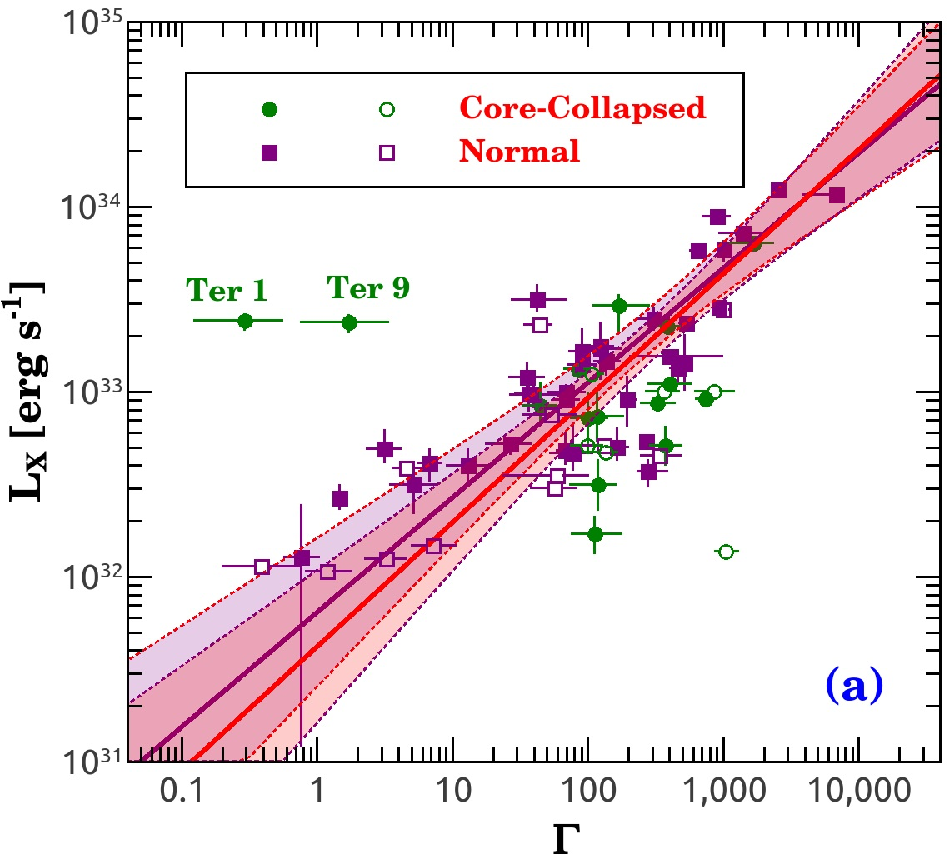}\includegraphics[width=0.5\linewidth]{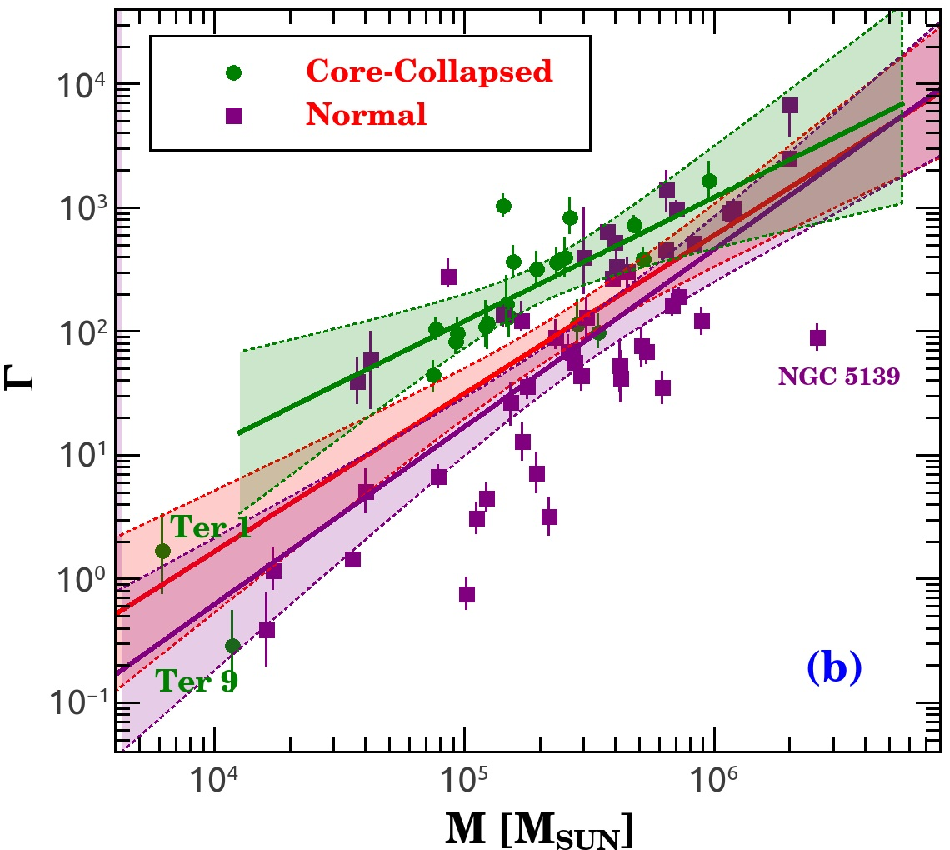}
\linespread{0.7}
\caption{(a) GC X-ray luminosity as a function of the stellar encounter rate. (b) The stellar encounter rate versus cluster mass $M$. Color-coded symbols denote the different types of GCs as in Figure~\ref{fig:X-K}. 
Solid lines and the associated shaded area are the power-law fitting functions and the $95\%$ confidence range, purple for the dynamically normal GCs, olive for the core-collapsed GCs and red for the full sample. The two outliers, Terzan 1 and Terzan 9, were not included in the fit. \label{fig:X-M-G}}
\end{figure*}   
 
To quantify the dynamical interactions in GCs, we adopt the updated stellar encounter rates by \citet{bahramian2013}. These authors have reconstructed the stellar density profiles of GCs based on their surface brightness profiles, thus the dynamically normal and core-collapsed GCs can be treated equally. 
More importantly, with Monte-Carlo simulations, errors in $\Gamma$ can also be properly estimated from the uncertainty in the observables such as distance, reddening and surface brightness. 
Specifically, we adopt $\Gamma_{1}\propto 4\pi\sigma_{c}^{-1}\int \rho^{2}(r)r^{2}dr$ from Table 4 of \citet{bahramian2013}, which is integrated over the entire cluster and normalized to a value of 1000 for NGC\,104, and is compatible to our measurement of the X-ray luminosity. 

In Figure~\ref{fig:X-M-G}a, we plot $L_{X}$ as a function of $\Gamma$ for each GC. It is evident that a strong correlation between $L_{X}$ and $\Gamma$ exists for the dynamically normal GCs. 
The Spearman's rank correlation coefficient is $r=0.769 \pm 0.093$, with the p-value of $p=1.7 \times 10^{-10}$ for random correlation. This correlation is still significant when taking the core-collapsed GCs into account, with $r=0.627 \pm 0.078$ and $p=1.4 \times 10^{-8}$ for the total GCs. 
The best fitting functions for the dynamically normal and total GCs can be written as $L_{X} \propto \Gamma^{0.62 \pm 0.06}$ and $L_{X} \propto \Gamma^{0.67 \pm 0.07}$, respectively (purple and red lines in Figure~\ref{fig:X-M-G}a). 
These relations are consistent with the finding of \citet{pooley2003} that, above a limiting luminosity of $4\times10^{30} \rm\ erg\ s^{-1}$, the number of detected X-ray sources in 12 GCs is proportional to the encounter rate, with $N_{X} \propto \Gamma^{0.74 \pm 0.36}$. 
A similar result has been obtained by \citet{maxwell2012}, with $N_{X} \propto \Gamma^{0.55 \pm 0.09}$ again based on only 12 GCs. 

The core-collapsed GCs show no significant correlation among themselves, but notably many of them are located below the best-fitting correlation for the dynamically normal GCs. 
This is contrary to the work of \citet{pooley2003}, \citet{lugger2007} and \citet{maxwell2012}, in which most core-collapsed GCs appeared abundant in X-ray sources and were located above the best-fitting correlation in their $N_{X}-\Gamma$ diagram. 
We suggest that this discrepancy is most likely due to their different adoption of $\Gamma$, which, based on the tranditional method, might have been underestimated for core-collapsed GCs.
This is hinted in Figure~\ref{fig:X-M-G}b, in which $\Gamma$ from \citet{bahramian2013} is plotted against $M$. Evidently, most core-collapsed GCs have a larger $\Gamma$ than the dynamically normal GCs at a given cluster mass. 
With this updated $\Gamma$, \citet{bahramian2013} first noted the lower abundance of X-ray sources in core-collapsed GCs than in their dynamically normal counterparts, which is consistent with our results in Figure~\ref{fig:X-M-G}a. 

The two significant outliers in the $L_{X}-\Gamma$ diagram, Terzan 1 and Terzan 9, deserve some remarks. Both show an X-ray luminosity significantly higher than expected from their $\Gamma$. 
\citet{cackett2006} detected 14 X-ray sources in the central 1.4 arcmin of Terzan 1, about 20 times larger than expected from the $N_{X}-\Gamma$ relation of \citet{pooley2003}. 
For Terzan 9, with the 15.2 ks {\it Chandra observation}, we detected 4 X-ray sources within the half-light radius, also more than expected from the $N_{X}-\Gamma$ relation\footnote{The faintest X-ray source in Terzan 9 has a luminosity of $\sim$ $4\times10^{31} \rm\ erg\ s^{-1}$. 
For its encounter rate of $1.71^{+1.67}_{-0.959}$ \citep{bahramian2013}, only $\lesssim1$ source with a luminosity greater than $4\times10^{30}\rm\ erg\ s^{-1}$ is predicted by the $N_{X}-\Gamma$ relation of \citet{pooley2003}}. 
As noted in Section~\ref{subsec:2MASS}, at face value the K-band luminosities of these two clusters are not compatible with their V-band magnitudes, which are likely due to strong foreground extinction. Therefore $\Gamma$, evaluated based on V-band data in \citet{bahramian2013}, might have been underestimated for both clusters, 
and it is premature to claim an over-abundance of X-ray sources in these two clusters. 

\begin{figure*}[htbp]
\centering
\includegraphics[width=0.5\textwidth]{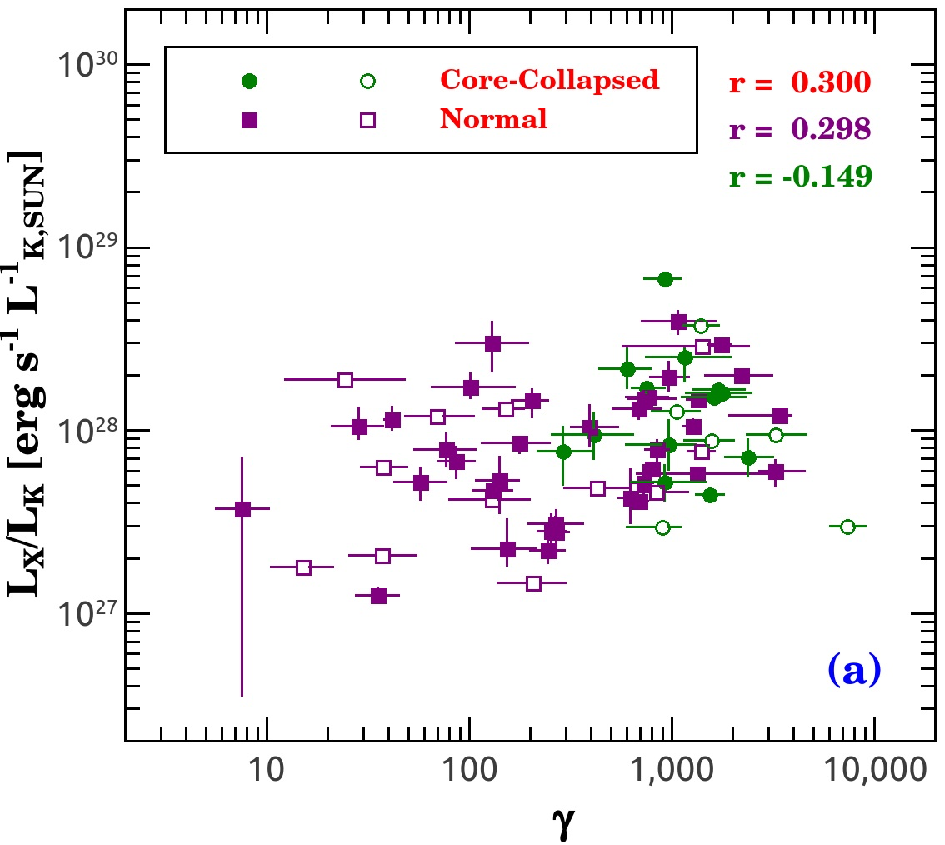}\includegraphics[width=0.5\textwidth]{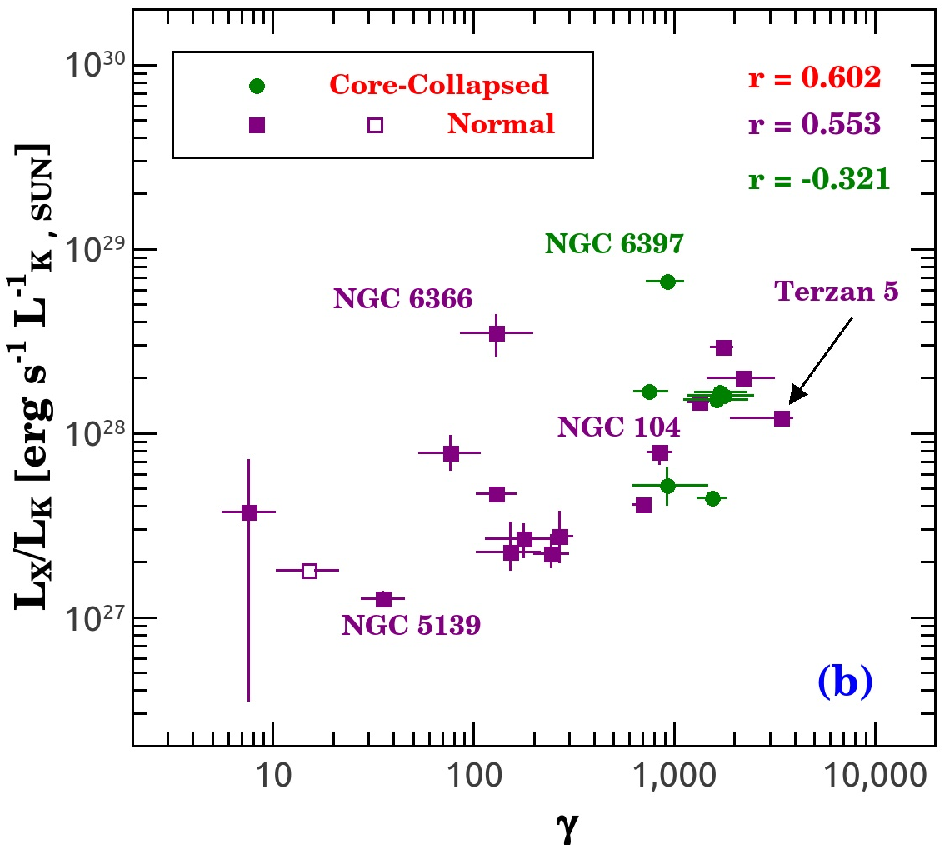}
\linespread{0.7}
\caption{$L_{X}/L_{K}$ as a function of the specific encounter rate $\gamma$ for (a) all the 69 GCs cosidered in this work, and (b) the 23 GCs considered in \citet{pooley2006}. 
Color coded symbols denote different types of GCs as in Figure~\ref{fig:X-K}. Some well-studied GCs are marked with their names. \label{fig:LXLK}}
\end{figure*}

Now it becomes clear that $L_{X}$ depends on both $M$ and $\Gamma$, and so does $N_{X}$, the number of detected X-ray sources in a given GC. 
To minimize the dependence on cluster mass, \citet{pooley2006} and \citet{pooley2010} have studied the specific number of detected sources ($n_{x} \equiv N_{X}/M_{6}$) as a function of the specific encounter rate ($\gamma \equiv \Gamma/M_{6}$), where $M_{6}$ is the cluster mass in units of $10^{6}{\rm M_{\odot}}$. 
\citet{pooley2006} found that the two parameters are correlated, with a fitting function $n_{x} = C+A \gamma^{\alpha}$, where $C$, $A$, and $\alpha$ are free parameters. In this regard, the origin of X-ray sources in GCs can be attributed to two channels: descendants of primordial binaries ($C$) and a dynamically formed population ($A \gamma^{\alpha}$). 

Similarly, we plot $L_{X}/L_{K}$ versus $\gamma$ in Figure~\ref{fig:LXLK}, showing all 69 GCs in Figure~\ref{fig:LXLK}a, but only those 23 GCs considered by \citet{pooley2006} in Figure~\ref{fig:LXLK}b to facilitate a direct comparision.
Since $\Gamma$ were estimated based on the V-band data by \citet{bahramian2013}, here we have calculated $\gamma$ ($\gamma \equiv \Gamma/M_{6}$) using the V-band-based mass (Figure~\ref{fig:K-M}) for consistency.
Both Figure 2d of \citet{pooley2006} and our Figure~\ref{fig:LXLK}b show a moderate increase of X-ray source abundance (in terms of $n_{x}$ or $L_{X}/L_{K}$) with increasing $\gamma$\footnote{Besides the aforementioned difference in the estimation of $\gamma$ for core-collapse GCs between the two diagrams, Terzan 5 in Figure 2d of \citet{pooley2006} shows a very high specific number of X-ray sources, which we suggest is overestimated because the mass of Terzan 5 calculated from its V-band magnitude is likely an underestimate.}. 
However, the much larger sample in Figure~\ref{fig:LXLK}a does not support a significant correlation;
the Spearman's rank correlation coefficients for the dynamically normal, core-collapse and total GCs are $r=0.298 \pm 0.092$, $r=-0.149\pm 0.149$ and $r=0.300 \pm 0.078$, with random correlation p-value of $p=0.040$, $p=0.542$ and $p=0.013$, respectively. 

We end this subsection by emphasizing that a significant correlation exists between $\Gamma$ and $M$ (Figure~\ref{fig:X-M-G}b). The Spearman's rank coefficients for the dynamically normal, core-collapsed and total GCs are $r = 0.754 \pm 0.092$, $0.660 \pm 0.149$ and $0.611 \pm 0.078$, respectively, with the p-value of $5.365 \times 10^{-11}$, $0.002$ and $8.543 \times10^{-8}$ for random correlation.
We fit the $\Gamma-M$ correlation with a power law function, which gives $\Gamma \propto M^{1.43 \pm 0.17}$ for dynamically normal GCs, $\Gamma \propto M^{1.01 \pm 0.26}$ for core-collapsed GCs, and $\Gamma \propto M^{1.28 \pm 0.17}$ for total GCs. 
The fitting functions are also plotted in Figure~\ref{fig:X-M-G}b.

\subsection{Correlations with other physical parameters}

Observationally, LMXBs are more likely to be found in old, metal-rich GCs, rather than in young, metal-poor ones \citep{bellazzini1995,kundu2002,sarazin2003,kim2006,sivakoff2007,li2010,paolillo2011,kim2013}. 
This indicates that the formation and evolution of LMXBs in GCs are affected by stellar metallicity. 
It has been suggested that a smaller convection zone in metal-poor stars relative to metal-rich stars at a given mass may help turn off the magnetic braking during binary evolution, leading to a lower formation efficiency of LMXBs in metal-poor GCs \citep{ivanova2006}. 
Alternatively, a stronger irradiation-induced wind from metal-poor stars, due to less efficient line cooling and energy dissipation, will speed up the evolution of LMXBs in the host cluster and reduce their number \citep{maccarone2004}. 

\begin{figure*}[htbp]
\centering
\includegraphics[width=1.0\textwidth]{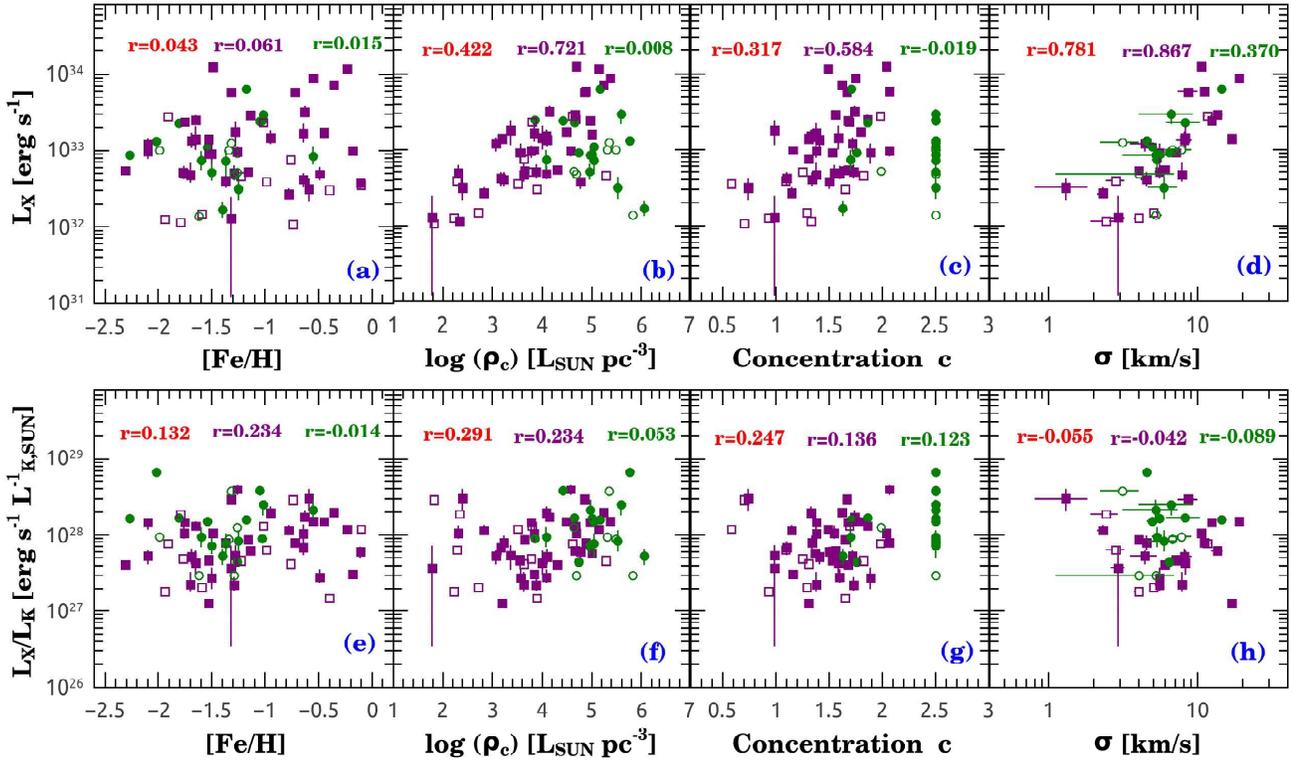}
\linespread{0.7}
\caption{Dependence of the X-ray luminosity (top panels) and $L_{X}/L_{K}$, a proxy of the X-ray emissivity (bottom panels), on various physical properties of GCs. 
From left to right: Metallicity, central luminosity density, King model central concentaration and central velocity dispersion. 
All these parameters are adopted from \citet{harris1996}. 
Color-coded symbols represent the different types of GCs as in Figure~\ref{fig:X-K}, and same color-coded text give the Speraman rank correlation coefficients: olive for core-collapsed, purple for dynamically normal and red for the total. \label{fig:cor1}}
\end{figure*}

In Figure~\ref{fig:cor1}a and \ref{fig:cor1}e, we test the dependence of $L_{X}$ and $L_{X}/L_{K}$ on metallicity. No clear correlation exists for these quantities.
Presumably the same physical processes such as magnetic braking and irradiation also happen in CVs and ABs, but the influence of metallicity in these processes appears less important than in the case of LMXBs. 

In Figure \ref{fig:cor1}b, we study the dependence of $L_{X}$ on the cluster central luminosity density $\rho_{c}$. It is evident that the dynamically normal GCs have a larger $L_{X}$ with increasing $\rho_{c}$.
The core-collapsed GCs have higher central stellar densities, but they exhibit no similar correlation between $L_{X}$ and $\rho_{c}$. 
On the other hand, dependence of $L_{X}/L_{K}$ on $\rho_{c}$ is not evident in Figure \ref{fig:cor1}f for either the dynamically normal or core-collapsed GCs. This suggests that cluster mass is the more fundamental parameter underlying the $L_{X}-\rho_{c}$ relation.

In Figure \ref{fig:cor1}c and \ref{fig:cor1}d, a positive correlation exists between $L_{X}$ and the cluster central concentration $c$ or the central velocity dispersion $\sigma$. We note that similar relations between GC V-band absolute magnitude and $c$ or $\sigma$ have been found by \citet{djorgovski1994}. 
These two correlations may again reflect the more fundamental dependency on cluster mass. Indeed, Figure \ref{fig:cor1}g and \ref{fig:cor1}h show no significant correlation between $L_{X}/L_{K}$ and $c$ or between $L_{X}/L_{K}$ and $\sigma$.

\begin{figure*}[htbp]
\centering
\includegraphics[width=1.0\textwidth]{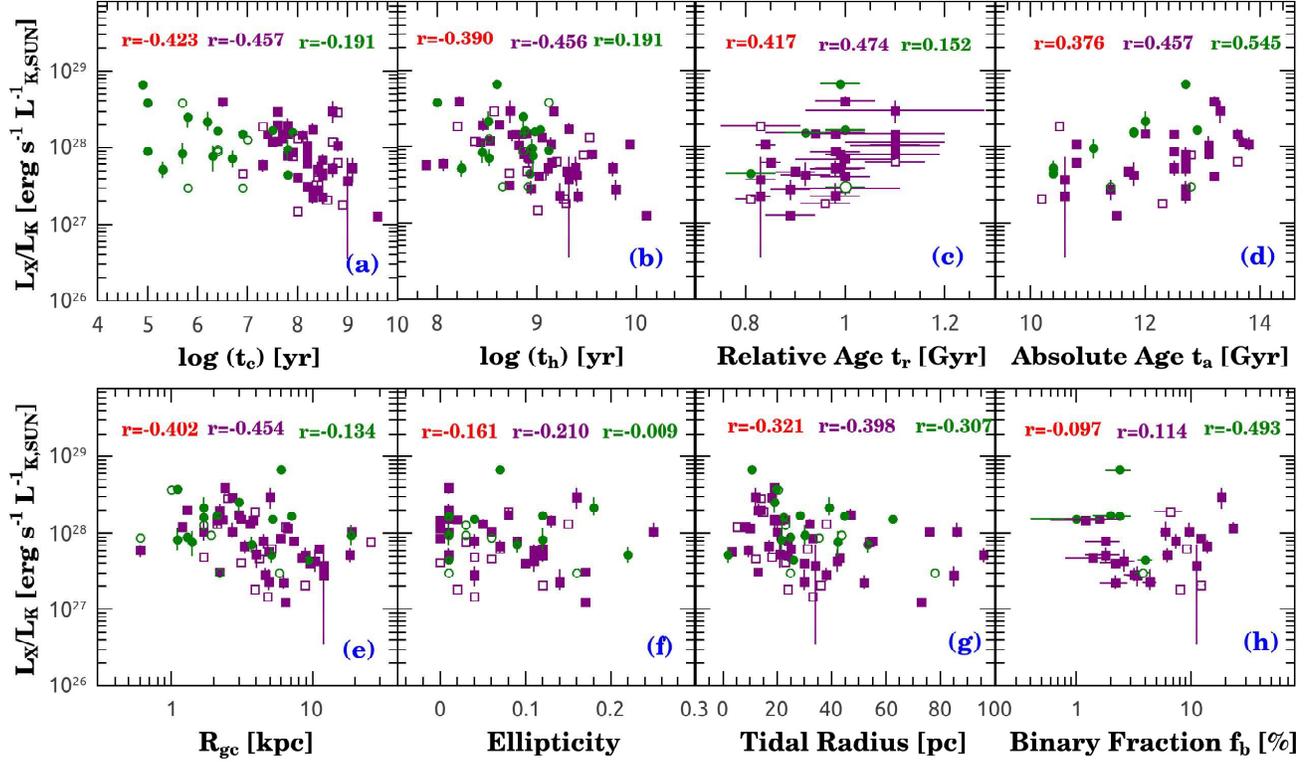}
\linespread{0.7}
\caption{Dependence of $L_{X}/L_{K}$, a proxy of the X-ray emissivity on various global properties of GCs. Top panels from left to right: Cluster core relaxation time, median relaxation time, relative age and absolute age. 
Bottom panels from left to right: Cluster distance to Galactic center, ellipticity of optical isophotes, tidal radius and main sequence binary fraction. 
All these parameters are adopted from \citet{harris1996}, except that the cluster relative age is from \citet{Marín-Franch2009}, absolute age from \citet{forbes2010} and the main sequence binary fraction from \citet{milone2012}. 
Color-coded symbols and text have the same meanings as in Figure \ref{fig:cor1}. \label{fig:cor2}}
\end{figure*}

In Figure~\ref{fig:cor2}a-d, we plot $L_{X}/L_{K}$ as a function of GC core relaxation time $t_{c}$, median relaxation time $t_{h}$, relative age $t_{r}$ and absolute age $t_{a}$. 
In general, GCs with smaller $t_{c}$ or $t_{h}$ are dynamically older. It appears that the GC X-ray emissivity increases with the dynamical age, as suggested by a marginally significant negative dependence of $L_{X}/L_{K}$ on $t_{c}$ and $t_{h}$ (Figure~\ref{fig:cor2}a and \ref{fig:cor2}b, respectively).
The marginal dependence of $L_{X}/L_{K}$ on the relative age $t_{r}$ or the absolute age $t_{a}$ (Figure~\ref{fig:cor2}c and \ref{fig:cor2}d) is also consistent with such a mild trend. 

In Figure~\ref{fig:cor2}e-g, we plot $L_{X}/L_{K}$ as a function of the GC distance to the Galactic center $R_{gc}$, ellipticity of optical isophotes $e$, tidal radius $r_t$ and main sequence binary fraction $f_b$. 
In principle, with increasing distances away from the Galactic center, GCs will suffer from weaker tidal force and grow their tidal radius, affecting its dynamical and geometrical structure and potentially the abundance of binaries. 
This is suggested by the mild anti-correlations in Figure~\ref{fig:cor2}e and \ref{fig:cor2}g. 
However, we find no statistically significant correlation between $L_{X}/L_{K}$ and ellipticity of optical isophotes $e$ in Figure~\ref{fig:cor2}f. Although the main sequence binary fraction ($f_{b}$) may be closely related to the abundance of the weak X-ray sources, Figure~\ref{fig:cor2}h shows no significant correlation between $L_{X}/L_{K}$ and $f_{b}$.

To conclude this section, we summarize in Table~\ref{tab:spearman} the Spearman's rank coefficients for all the tested correlations. 

\section{Discussion}

It has been known for over 40 years that the abundance of luminous LMXBs in GCs exceeds that of the field by orders of magnitude, which is generally accepted as the consequence of efficient dynamical formation of neutron star binaries in the dense environment of GCs. 
Hut \& Verbunt (1983) were among the first to predict that dynamical processes would lead to the formation of as many white dwarf binaries as neutron star binaries in GCs. 
Pooley and colleagues, upon finding a correlation between the number of weak X-ray sources and the stellar encounter rate for a moderate sample of GCs, argued that the X-ray populations, in particular CVs, are over-abundant in GCs. 
Our analysis in Section 3, however, points to an opposite trend: most GCs exhibit a lower cumulative X-ray emissivity than found in the field, which is represented by the Solar neighborhood and the Local Group dwarf elliptical galaxies. 
Because in all these environments the cumulative X-ray emissivity is a reasonable proxy of the source abundance, the immediate conclusion is that the weak X-ray populations, primarily CVs and ABs, are under-abundant in GCs as compared to the field.
In the following, we address the implications of this result.

\subsection{Dependence on stellar encounter rate}
The under-abundance of weak X-ray sources in GCs demands for a revisit of the $N_X-\Gamma$ \citep{pooley2003} and $L_X-\Gamma$ (Figure~\ref{fig:X-M-G}a) correlations as evidence for a predominantly dynamical origin of the X-ray populations. 
As shown in Figure~\ref{fig:X-M-G}b, the stellar encounter rate $\Gamma$ is strongly correlated with the cluster mass $M$, thus it is not unreasonable to raise the question of whether cluster mass is the more fundamental parameter underlying the $N_X-\Gamma$ and $L_X-\Gamma$ relations.   
Indeed, in Section 3.1 we have demonstrated a correlation between $L_X$ and $L_K$ (hence $M$), which is statistically as significant as the $L_X-\Gamma$ relation, according to the Spearman's rank cofficients. 
Moreover, our larger GC sample disfavors a strong positive correlation between the source abundance (traced by $L_X/L_K$) and the specific encounter rate, $\gamma$ (Figure~\ref{fig:LXLK}a), as previously suggested by \citet{pooley2006}. 

\begin{figure}[htbp]
\centering
\includegraphics[width=.5\textwidth]{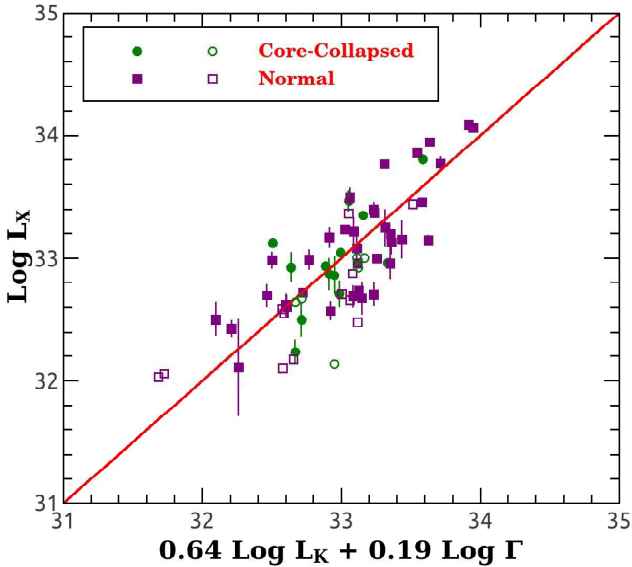}
\linespread{0.7}
\caption{GC X-ray luminosity expressed as a function of K-band luminosity and stellar encounter rate:  
The olive dots and purple squares donate the core-collapsed and dynamically normal GCs, respectively. \label{fig:X-K-G}}
\end{figure}

To further compare the influence of the cluster mass and stellar encounters on the X-ray populations, we test a correlation between $L_X$, $M$ and $\Gamma$ using the following form: ${\rm log}\ L_{X} = a_K\,{\rm log}\ L_{K} + a_\Gamma\,{\rm log}\ \Gamma + c$. 
Giving equal weights to $L_K$ and $\Gamma$ in the regression, we find $a_K=0.64 \pm0.12$, $a_\Gamma=0.19\pm0.07$ and $c=29.4\pm0.52$ (Figure~\ref{fig:X-K-G}). 
It appears that the cluster mass, rather than the stellar encounter rate, is the predominant factor in determining the amount of weak X-ray sources. 
This implies that a substantial fraction of the X-ray populations, in particular CVs and ABs, are descendants of primordial binaries, although dynamical processes must have played at least a partial role in the formation of the X-ray populations.
In any case, taking the $N_X-\Gamma$ and $L_X-\Gamma$ relations as solid evidence for a dynamical origin of the X-ray populations may be an over-simplification. Below, we argue that not all dynamical interactions in GCs will lead to the formation of X-ray sources. 

\subsection{Efficiency of the primordial channel of close binary formation}

In general, when considering the formation of close binary systems in a dense stellar environment such as GCs, three competing and coupled effects are relevant.
The first is normal stellar evolution that eventually turns wide, primordial binaries into close binaries, which is referred to as the {\it primordial channel}. 
The second is the {\it dynamical formation} of close binaries due to two-body or three-body interactions.
The third and a negative effect is the {\it dynamical disruption} of primordial binaries mainly due to three-body (binary-single) or four-body (binary-binary) interactions \citep{heggie1975,hills1975,mikkola1983,hut1992a,hut1992b,hut1993,bacon1996}.
For LMXBs, because of the rarity of neutron stars and the relatively short lifetime of their progenitor stars, the first and third effects are generally negligible.
However, for CVs and ABs, because their progenitors are mainly low-mass stars, they should evolve on a timescale comparable to or even greater than the GC relaxation time, making all the above three effects relevant.
Therefore, we argue that the under-abundance of CVs and ABs in GCs reflects a low efficiency of the primordial channel, i.e., a substantial fraction of the primordial binaries are dynamically disrupted before they can otherwise evolve into close binaries. 
Such a scenario is supported by both theoretical and observational studies.

Theoretically, the fate of binaries in GCs is governed by both normal stellar evolution and stellar dynamical interactions.
In the former case, a number of stellar processes, such as Roche lobe msss tranfer, steller wind and supernova kick, can substantially modify the binary oribits and even lead to orbit disruption or binary merger.
Dynamical interactions can alter binaries in a more abrupt manner. 
For example, most ``soft" binaries in GCs will be destroyed if they suffer from a strong encounter with other stars. 
Even for the ``hard" binaries, their interactions with other stars can exchange one of the primordial members with the intruding star, in the meantime causing the orbit to expand or shrink, modifying the orbit eccentricity, enhancing the systemic velocity via gravitational recoil, or leading to physical collision. 
Numerical simulations have confirmed that binaries can be dynamically disrupted in GCs \citep{ivanova2005,fregeau2009,chatterjee2010}. 
The probability for a primordial binary to survive from dynamical destruction depends on the binary hardness and the collision timescale in GCs. 
Comparing with simulations that only consider normal stellar evolution, simulations taking into account binary destruction result in a much lower binary fraction in GCs \citep{ivanova2011}. 

\begin{figure}[htbp]
\centering
\includegraphics[width=0.5\textwidth]{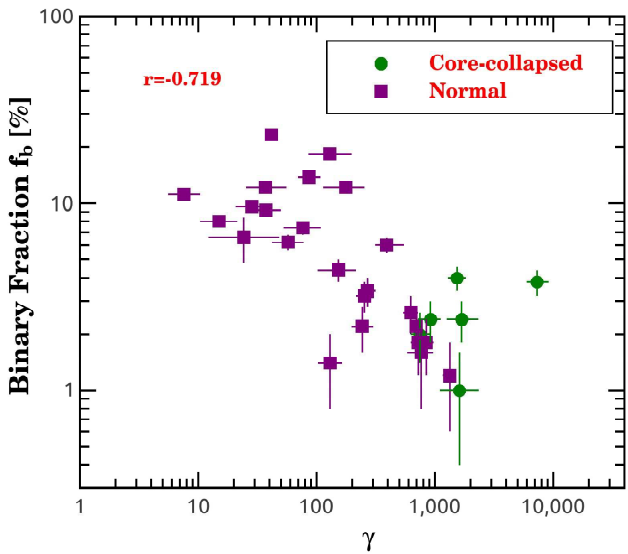}
\linespread{0.7}
\caption{Main sequence binary fraction of GCs \citep{milone2012} as a function of the specific encounter rate $\gamma$. The olive dots and purple squares donate the core-collapsed and normal GCs, respectively. The Spearman's rank correlation coefficient is marked in red text. \label{fig:fb}}
\end{figure}

Observationally, the main sequence binary fraction is found to be on average much lower in GCs than in the Solar neighborhood \citep{Cote1996,albrow2001,davis2008} or open clusters \citep{sollima2010}. 
Recent numerical simulations suggest that these present-day binary fractions are consistent with a universal, near-unity initial binary fraction in GCs \citep{leigh2015}. 
If this were the case, dynamical disruption of primordial binaries must have been efficient. This is also suggested by the observed trend that the main sequence binary fraction of GCs decreases with increasing cluster age \citep{sollima2007,milone2012,ji2015}.
Furthemore, as shown in Figure~\ref{fig:fb}, a negative correlation is evident between the main sequence binary fraction and the specific stellar encounter rate, strongly suggesting dynamical disruption of primordial binaries and consequently the dearth of X-ray-emitting close binaries.

\subsection{Dynamical formation versus dynamical disruption}
 
Nevertheless, dynamical formation of CVs and ABs must be taking place in GCs. 
In reality, it is most likely that dynamical formation and destruction of binaries occur simultaneously and persistently in GCs. 
Fortunately, the two competing effects may be unified in one process, namely, the binary-single interaction. 
In general, such interactions obey the Hills-Heggie law, which states that hard binaries (with $|E_{b}| > E_{k}$) evolve into smaller orbits, while soft binaries (with $|E_{b}| < E_{k}$) tend to be softer and even disrupted \citep{hills1975,heggie1975}.  

More quantitatively, we may express the abundance of X-ray binaries as, 
\begin{equation}
L_{X}/L_K \approx N_X \bar{L}_{X} / (N_{*} \bar{L}_{K})
= (N_b/N_{*})(N_X/N_b)\bar{L}_{X}/\bar{L}_{K},
\end{equation}
where $\bar{L}_{X}$ ($\bar{L}_{K}$)
is the characteristic X-ray (K-band) luminosity of a binary (star), $N_b/N_{*} \approx f_b$ is the GC binary fraction, and $N_X/N_b$ is the fraction of binaries being an X-ray-emitting close binary.  
Recall that the average GC X-ray emissivity is $\sim$40\% lower than that of the cumulative X-ray emissivity of the field populations ($6.9\times10^{27}\rm\ erg\,s^{-1}M^{-1}_{\odot}$ vs. $11.8\times10^{27}\rm\ erg\,s^{-1}M^{-1}_{\odot}$; Section 3.1).
On the other hand, the main sequence binary fraction in the Solar neighborhood is $\sim$40\% \citep{fischer1992}, while most measurements of GC binary fraction range between 1-20\% (Figure 7h). 

If we assume that GCs have an initial binary fraction at least comparable to that of the Solar neighborhood -- in fact, a near-unity initial binary fraction has been suggested for GCs (Leigh et al.~2015) -- and that the Solar neighborhood binaries had evolved little due to the lack of dynamical interactions,
from the above we may infer that the reduction in binary fraction (by a factor of 2-40 in $N_b/N_{*}$) due to dynamical disruption of ``soft'' primordial binaries has been partially compensated by the dynamical formation of X-ray-emitting ``hard'' binaries (i.e., an enhanced $N_X/N_b$ as compared to the field). 
The statistical outcome of this competition appears to be a mild anti-correlation between the abundance of the X-ray sources and the cluster mass, $\epsilon_{X} \propto M^{-0.31 \pm 0.09}$ (Section 3.1), 
which may be qualitatively understood in the sense that dynamical disruption of primordial binaries is more efficient in more massive GCs, because of their higher encounter rate ($\Gamma \propto M^{1.28}$ in Figure~\ref{fig:X-M-G}b) and higher stellar velocity dispersion ($E_{k} \propto \sigma^{2}$). 
On the other hand, the competition may allow the abundance of weak X-ray sources in some dynamically old GCs to catch up with, or even exceeds, the field level. For example, NGC 6397 has the highest X-ray emissivity among all GCs (Figures~\ref{fig:em} and 5), and the binary fraction in this cluster is also very low ($f_b=2.4\pm0.6\%$, \citealt{milone2012}), which suggests that the compensation of weak X-ray sources by dynamical formation channel have surpassed the dynamical disruption channel, and many ``hard" main sequence binaries have been dynamically transformed into weak X-ray sources in this cluster.
We leave a more quantitative study of the effect of binary-single interaction on the X-ray source abundance to a future work.  
 
\section{Summary}

In this work, we have surveyed the cumulative X-ray emission of the largest-so-far sample (69) of Milky Way globular clusters observed by {\it Chandra}. Our main findings are as follows.

1. The X-ray emissivity of most GCs is lower than that of the Solar neighborhood and dwarf elliptical galaxies, indicating that a paucity of weak X-ray sources, mainly CVs and ABs, in GCs with respect to the field populations. 
This under-abundance of X-ray-emitting close binaries suggests that formation of such systems through the primordial channel is suppressed in GCs, which is likely due to dynamical disruption of primordial binaries before they can evolve into CVs and ABs. 

2. The GC X-ray luminosity, stellar encounter rate and cluster mass are highly correlated among each other, with $L_{X} \propto \Gamma^{0.67 \pm 0.07}$, $L_{X} \propto M^{0.74 \pm 0.13}$ and $\Gamma \propto M^{1.28 \pm 0.17}$. 
Furthermore, the GC X-ray luminosity can be expressed as a function of cluster mass and stellar encounter rate, with ${\rm Log}\ L_{X}=(29.40\pm0.47) + (0.64\pm0.12)\ {\rm Log}\ L_{K} + (0.19\pm0.07)\ {\rm Log}\ \Gamma$, 
which suggests that dynamical formation of CVs and ABs in GCs is less dominant than previously thought, while the primordial channel may still have a substantial contribution. 

3. Binary-single interactions, which would lead to both the disruption of soft primordial binaries and the formation of close binaries, seem to provide a natural explanation of the above results. The net outcome is the present-day abundance of weak X-ray sources in GCs, which appears no higher than that of the field.

4. Correlations between the GC X-ray emissivity and global cluster properties such as metallicity, dynamical age and structural parameters have been examined, but no statistically significant trends can be revealed.

\acknowledgements
We thank the anonymous referee for valuable comments that help improve our manuscript. This work is supported by the National Natural Science Foundation of China under grants 11133001, 11333004 and 11303015, and the Strategic Priority Research Program of CAS under grant XDB09000000. Z.L. acknowledges support from the Recruitment Program of Global Youth Experts.

\begin{deluxetable}{lrrrrrrrrrrrrrr}
\tabletypesize{\scriptsize}
\tablecolumns{15}
\linespread{0.75}
\tablewidth{0pc}
\tablecaption{BASIC PROPERTIES OF GLOBULAR CLUSTERS}
\tablehead{
\colhead{Name} & \colhead{M} & \colhead{$\Gamma$} & \colhead{$-\delta$} & \colhead{$+\delta$} & \colhead{$\gamma$} & \colhead{$-\delta$} & \colhead{$+\delta$} & \colhead{D} & \colhead{E(B-V)} & \colhead{$r_h$} & \colhead{$r_h/r_c$} & \colhead{Fe/H} & \colhead{$\rho_{c}$} & \colhead{$t$}\\
\colhead{(1)} & \colhead{(2)} & \colhead{(3)} & \colhead{(4)} & \colhead{(5)} & \colhead{(6)} & \colhead{(7)} & \colhead{(8)} & \colhead{(9)} & \colhead{(10)} & \colhead{(11)} & \colhead{(12)} & \colhead{(13)} & \colhead{(14)} & \colhead{(15)}}
\startdata
\sidehead{\textbf{Normal GCs:}}
NGC 104   & 118.5 & 1000  & 134  & 154  & 844  & 113  & 130   & 4.5  & 0.04 & 3.17  & 8.8  & -0.72 & 4.88 & 13.06 \\
NGC 288   & 10.13 & 0.766 & 0.205& 0.284& 7.56 & 2.02 & 2.80  & 8.9  & 0.03 & 2.23  & 1.7  & -1.32 & 1.78 & 10.62 \\
NGC 2808  & 115.2 & 923   & 82.7 & 67.2 & 801  & 71.8 & 58.3  & 9.6  & 0.22 & 0.8   & 3.2  & -1.14 & 4.66 & 10.8  \\
NGC 3201  & 19.3  & 7.17  & 2.27 & 3.56 & 37.1 & 11.8 & 18.4  & 4.9  & 0.24 & 3.1   & 2.4  & -1.59 & 2.71 & 10.24 \\
NGC 5024  & 61.6  & 35.4  & 9.6  & 12.4 & 57.5 & 15.6 & 20.1  & 17.9 & 0.02 & 1.31  & 3.7  & -2.1  & 3.07 & 12.67 \\
NGC 5139  & 256.8 & 90.4  & 20.4 & 26.3 & 35.2 & 7.94 & 10.2  & 5.2  & 0.12 & 5     & 2.1  & -1.53 & 3.15 & 11.52 \\
NGC 5272  & 72.04 & 194   & 18   & 33.1 & 269  & 25   & 45.9  & 10.2 & 0.01 & 2.31  & 6.2  & -1.5  & 3.57 & 11.39 \\ 
NGC 5286  & 63.33 & 458   & 60.7 & 58.4 & 723  & 95.8 & 92.2  & 11.7 & 0.24 & 0.73  & 2.6  & -1.69 & 4.1  & 12.54 \\
NGC 5824  & 70.08 & 984   & 155  & 171  & 1400 & 221  & 244   & 32.1 & 0.13 & 0.45  & 7.5  & -1.91 & 4.61 & 12.8  \\
NGC 5904  & 67.55 & 164   & 30.4 & 38.6 & 243  & 45   & 57.1  & 7.5  & 0.03 & 1.77  & 4.0  & -1.29 & 3.88 & 10.62 \\
NGC 5927  & 26.89 & 68.2  & 10.3 & 12.7 & 254  & 38.3 & 47.2  & 7.7  & 0.45 & 1.1   & 2.6  & -0.49 & 4.09 & 12.67 \\
NGC 6093  & 39.59 & 532   & 68.8 & 59.1 & 1340 & 174  & 149   & 10   & 0.18 & 0.61  & 4.1  & -1.75 & 4.79 & 12.54 \\ 
NGC 6121  & 15.19 & 26.9  & 9.56 & 11.6 & 177  & 62.9 & 76.4  & 2.2  & 0.35 & 4.33  & 3.7  & -1.16 & 3.64 & 12.54 \\
NGC 6139  & 44.63 & 307   & 82.1 & 95.4 & 688  & 184  & 214   & 10.1 & 0.75 & 0.85  & 5.7  & -1.65 & 4.67 & --    \\
NGC 6144  & 11.11 & 3.14  & 0.85 & 1.07 & 28.3 & 7.65 & 9.63  & 8.9  & 0.36 & 1.63  & 1.7  & -1.76 & 2.31 & 13.82 \\
NGC 6205  & 53.16 & 68.9  & 14.6 & 18.1 & 130  & 27.5 & 34    & 7.1  & 0.02 & 1.69  & 2.7  & -1.53 & 3.55 & 11.65 \\
NGC 6218  & 16.97 & 13.0  & 4.03 & 5.44 & 76.6 & 23.8 & 32.1  & 4.8  & 0.19 & 1.77  & 2.2  & -1.37 & 3.23 & 12.67 \\
NGC 6287  & 17.77 & 36.3  & 7.74 & 7.70 & 204  & 43.6 & 43.3  & 9.4  & 0.6  & 0.74  & 2.6  & -2.1  & 3.78 & 13.57 \\  
NGC 6304  & 16.81 & 123   & 22   & 53.8 & 732  & 131  & 320   & 5.9  & 0.54 & 1.42  & 6.8  & -0.45 & 4.49 & 13.57 \\
NGC 6333  & 30.59 & 131   & 41.8 & 59.1 & 428  & 137  & 193   & 7.9  & 0.38 & 0.96  & 2.1  & -1.77 & 3.78 & --    \\
NGC 6341  & 38.87 & 270   & 29   & 30.1 & 695  & 74.6 & 77.4  & 8.3  & 0.02 & 1.02  & 3.9  & -2.31 & 4.3  & 13.18 \\
NGC 6352  & 7.83  & 6.74  & 1.3  & 1.71 & 86.1 & 16.6 & 21.8  & 5.6  & 0.22 & 2.05  & 2.5  & -0.64 & 3.17 & 12.67 \\
NGC 6362  & 12.18 & 4.56  & 1.03 & 1.51 & 37.4 & 8.46 & 12.4  & 7.6  & 0.09 & 2.05  & 1.8  & -0.99 & 2.29 & 13.57 \\
NGC 6366  & 3.996 & 5.14  & 1.76 & 2.75 & 129  & 44   & 68.8  & 3.5  & 0.71 & 2.92  & 1.3  & -0.59 & 2.39 & 13.31 \\
NGC 6388  & 117.4 & 899   & 213  & 238  & 766  & 181  & 203   & 9.9  & 0.37 & 0.52  & 4.3  & -0.55 & 5.37 & 12.03 \\
NGC 6401  & 29.21 & 44    & 10.7 & 11   & 151  & 36.6 & 37.7  & 10.6 & 0.72 & 1.91  & 7.6  & -1.02 & 3.79 & --    \\
NGC 6402  & 88.23 & 124   & 30.2 & 31.8 & 141  & 34.2 & 36    & 9.3  & 0.6  & 1.3   & 1.6  & -1.28 & 3.36 & --    \\
NGC 6440  & 63.91 & 1400  & 477  & 628  & 2193 & 746  & 983   & 8.5  & 1.07 & 0.48  & 3.4  & -0.36 & 5.24 & --    \\
NGC 6517  & 40.33 & 338   & 97.5 & 152  & 838  & 242  & 377   & 10.6 & 1.08 & 0.5   & 8.3  & -1.23 & 5.29 & --    \\
NGC 6528  & 8.58  & 278   & 49.5 & 114  & 3240 & 577  & 1330  & 7.9  & 0.54 & 0.38  & 2.9  & -0.11 & 4.77 & --    \\
NGC 6535  & 1.61  & 0.388 & 0.192& 0.389& 24.2 & 12   & 24.2  & 6.8  & 0.34 & 0.85  & 2.4  & -1.79 & 2.34 & 10.5  \\
NGC 6539  & 41.84 & 42.1  & 15.3 & 28.6 & 101  & 36.6 & 68.4  & 7.8  & 1.02 & 1.7   & 4.5  & -0.63 & 4.15 & --    \\
NGC 6553  & 25.92 & 69    & 18.8 & 26.8 & 266  & 72.5 & 103   & 6    & 0.63 & 1.03  & 1.9  & -0.18 & 3.84 & --    \\
NGC 6569  & 41.46 & 53.6  & 20.8 & 30.2 & 129  & 50.2 & 72.8  & 10.9 & 0.53 & 0.80  & 2.3  & -0.76 & 3.63 & --    \\
NGC 6626  & 37.12 & 648   & 91.1 & 83.8 & 1750 & 245  & 226   & 5.5  & 0.4  & 1.97  & 8.2  & -1.32 & 4.86 & --    \\
NGC 6637  & 22.99 & 89.9  & 18.1 & 36   & 391  & 78.7 & 157   & 8.8  & 0.18 & 0.84  & 2.5  & -0.64 & 3.84 & 13.06 \\
NGC 6638  & 14.24 & 137   & 27.1 & 38.6 & 962  & 190  & 271   & 9.4  & 0.41 & 0.51  & 2.3  & -0.95 & 4.09 & --    \\
NGC 6656  & 50.77 & 77.5  & 25.9 & 31.9 & 153  & 51   & 62.8  & 3.2  & 0.34 & 3.36  & 2.5  & -1.7  & 3.63 & 12.67 \\
NGC 6715  & 198.4 & 2520  & 274  & 226  & 1273 & 138  & 114   & 26.5 & 0.15 & 0.82  & 9.1  & -1.49 & 4.69 & 10.75 \\
NGC 6717  & 3.71  & 39.8  & 13.7 & 21.8 & 1070 & 369  & 587   & 7.1  & 0.22 & 0.68  & 8.5  & -1.26 & 4.58 & 13.18 \\
NGC 6760  & 27.64 & 56.9  & 19.4 & 26.6 & 206  & 70.2 & 96.2  & 7.4  & 0.77 & 1.27  & 3.7  & -0.4  & 3.89 & --    \\ 
NGC 6809  & 21.56 & 3.23  & 1    & 1.38 & 15   & 4.64 & 6.4   & 5.4  & 0.08 & 2.83  & 1.6  & -1.94 & 2.22 & 12.29 \\
NGC 6838  & 3.55  & 1.47  & 0.138& 0.146& 4.15 & 3.89 & 4.12  & 4    & 0.25 & 1.67  & 2.7  & -0.78 & 2.83 & 13.7  \\
NGC 7089  & 82.72 & 518   & 71.4 & 77.6 & 626  & 86.3 & 93.8  & 11.5 & 0.06 & 1.06  & 3.3  & -1.65 & 4    & 11.78 \\
Glim 01   & 30.00 & 400   & 200  & 200  & 8560 & 4280 & 12800 & 4.2  & 4.85 & 0.65  & 5.6  & --    & 5.0  & --    \\
Pal 10    & 4.18  & 59    & 35.5 & 42.8 & 1410 & 848  & 1020  & 5.9  & 1.66 & 0.99  & 1.2  & -0.1  & 3.51 & --    \\
Ter 3     & 1.71  & 1.18  & 0.37 & 0.65 & 68.9 & 21.4 & 38.1  & 8.2  & 0.73 & 1.25  &  1.1 & -0.74 & 1.82 & --    \\ 
Ter 5     & 200.0 & 6800  & 3020 & 1040 & 36200& 16100& 5540  & 5.9  & 2.38 & 0.72  & 3.4  & -0.23 & 5.14 & --    \\
\hline
\sidehead{\textbf{Core Collapse GCs:}}
NGC 362   & 47.6  & 735   & 117  & 137  & 1540 & 246  & 288  & 8.6  & 0.05 & 0.82  & 4.6  & -1.26 & 4.74 & 10.37 \\
NGC 1904  & 28.16 & 116   & 44.7 & 67.6 & 412  & 159  & 240  & 12.9 & 0.01 & 0.65  & 4.1  & -1.6  & 4.08 & 11.14 \\
NGC 5946  & 15.05 & 134   & 44.6 & 33.6 & 890  & 296  & 223  & 10.6 & 0.54 & 0.89  & 11.1 & -1.29 & 4.68 & 11.39 \\
NGC 6256  & 14.64 & 169   & 60.4 & 119  & 1150 & 413  & 813  & 10.3 & 1.09 & 0.86  & 43   & -1.02 & 5.59 & --    \\
NGC 6266  & 94.97 & 1670  & 569  & 709  & 1760 & 599  & 747  & 6.8  & 0.47 & 0.92  & 4.2  & -1.18 & 5.16 & 11.78 \\
NGC 6293  & 26.16 & 847   & 239  & 337  & 3240 & 914  & 1440 & 9.5  & 0.36 & 0.89  &17.8  & -1.99 & 5.31 & --    \\
NGC 6325  & 12.29 & 118   & 45.6 & 44.7 & 960  & 371  & 364  & 7.8  & 0.91 & 0.63  & 21   & -1.25 & 5.52 & --    \\
NGC 6342  & 7.48  & 44.8  & 12.5 & 14.4 & 599  & 167  & 193  & 8.5  & 0.46 & 0.73  & 14.6 & -0.55 & 4.97 & 12.03 \\
NGC 6355  & 34.17 & 99.2  & 25.7 & 41.1 & 290  & 75.2 & 120  & 9.2  & 0.77 & 0.88  & 17.6 & -1.37 & 5.04 & --    \\
NGC 6397  & 9.15  & 84.1  & 18.3 & 18.3 & 919  & 200  & 200  & 2.3  & 0.18 & 2.9   & 58   & -2.02 & 5.76 & 12.67 \\
NGC 6453  & 15.62 & 371   & 88.7 & 128  & 2380 & 568  & 820  & 11.6 & 0.64 & 0.44  & 8.8  & -1.5  & 4.95 & --    \\
NGC 6522  & 23.21 & 363   & 98.5 & 113  & 1560 & 424  & 478  & 7.7  & 0.48 & 1     & 20   & -1.34 & 5.48 & --    \\
NGC 6541  & 51.71 & 386   & 63.1 & 95.2 & 746  & 122  & 184  & 7.5  & 0.14 & 1.06  & 5.9  & -1.81 & 4.65 & 12.93 \\
NGC 6544  & 12.07 & 111   & 36.5 & 67.8 & 920  & 302  & 562  & 3    & 0.76 & 1.21  & 24.2 & -1.4  & 6.06 & 10.37 \\
NGC 6558  & 7.61  & 105   & 19.3 & 26.2 & 1380 & 253  & 344  & 7.4  & 0.44 & 2.15  & 71.7 & -1.32 & 5.35 & --    \\
NGC 6642  & 9.32  & 97.8  & 24.5 & 31.3 & 1050 & 263  & 336  & 8.1  & 0.4  & 0.73  & 7.3  & -1.26 & 4.64 & --    \\
NGC 6681  & 14.24 & 1040  & 192  & 267  & 7300 & 1350 & 1870 & 9    & 0.07 & 0.71  & 23.7 & -1.62 & 5.82 & 12.8  \\
NGC 6752  & 24.98 & 401   & 126  & 182  & 1610 & 504  & 729  & 4    & 0.04 & 1.91  & 11.24& -1.54 & 5.04 & 11.78 \\ 
NGC 7099  & 19.3  & 324   & 81.2 & 124  & 1680 & 421  & 642  & 8.1  & 0.03 & 1.03  & 17.2 & -2.27 & 5.01 & 12.93 \\
Ter 1     & 1.17  & 0.292 & 0.17 & 0.274& 24.9 & 14.5 & 23.3 & 6.7  & 1.99 & 1.4   & 95.5 & -1.03 & 3.85 & --    \\ 
Ter 9     & 0.616 & 1.71  & 0.959& 1.67 & 278  & 256  & 271  & 7.1  & 1.76 & 0.78  & 26   & -1.05 & 4.42 & --    \\
\enddata
\vspace{-0.3cm}
\tablecomments{Column 1-2: Name, total globular cluster mass (in units of $10^{4}M_{\odot}$) derived from V-band magnitude. Column 3-5: Stellar encounter rate from \citet{bahramian2013}, except for Glimpse 01, which is adopted from \citet{pooley2007}. $\pm\delta$ is the $1\sigma$ upper and lower limits of $\Gamma$. Column 6-8: specific encounter rate, which is calculated with $\gamma=\Gamma/M_{6}$, where $M_{6}$ is the GC mass in units of $10^{6}M_{\odot}$. Column 9-10: GC distance in units of kpc, foreground color excess. Column 11-12: half-light radius in arcmins, and ratio of half-light radius to core radius. Column 13-15: metallicity, central luminosity density in units of $log_{10}(L_{\odot}\rm\ pc^{-3})$ and absolute age of GCs in units of $\rm Gyr$ \citep{forbes2010}. }
\label{tab:GC}
\end{deluxetable}

\begin{deluxetable}{llllll}
\tabletypesize{\scriptsize}
\tablecolumns{6}
\tablecaption{OBSERVATION LOG}
\tablehead{
\colhead{GC Name} & \colhead{Chandra Obs.ID} & \colhead{Effect. exposure} & \colhead{Background} & \colhead{Cluster region} & \colhead{Source to Background} \\
\colhead{} & \colhead{} & \colhead{$t_{exp}$(ks)} & \colhead{Region} & \colhead{factor $k_{1}$} & \colhead{region factor $k_{2}$}\\
\colhead{(1)} & \colhead{(2)} & \colhead{(3)} & \colhead{(4)} & \colhead{(5)} & \colhead{(6)}}
\startdata
\sidehead{\textbf{Normal GCs:}}
NGC 104  & 953, 955          & 30.7, 30.7       & 2, 2    & 1.21, 1.16       & 0.249, 0.252        \\
NGC 288  & 3777              & 45.6             & 1       & 1.04             & 0.471               \\
NGC 2808 & 7453, 8560        & 45.4, 10.8       & 2, 2    & 1.71, 1.71       & 0.148, 0.147        \\
NGC 3201 & 11031             & 82.9             & 1       & 1                & 0.894               \\
NGC 5024 & 6560              & 24.2             & 2       & 1                & 0.316               \\
NGC 5139 & 13726, 13727      & 171, 47.6        & 1, 1    & 1.11, 1.10       & 0.440, 0.415        \\
NGC 5272 & 4542, 4543, 4544  & 9.9, 9.8, 9.4    & 1, 1, 1 & 1.05, 1.05, 1.05 & 0.942, 0.890, 0.914 \\
NGC 5286 & 8964, 9852        & 10, 3.3          & 2, 2    & 1, 1             & 0.232, 0.235        \\
NGC 5824 & 9026              & 10.7             & 2       & 1                & 0.200               \\
NGC 5904 & 2676              & 41.8             & 2       & 1                & 0.412               \\
NGC 5927 & 13673, 8953       & 46.5, 0.77       & 2, 2    & 1, 1             & 0.276, 0.290        \\
NGC 6093 & 1007              & 36.1             & 2       & 1                & 0.200               \\
NGC 6121 & 946, 7446, 7447   & 25.2, 47.9, 44.8 & 1, 1, 1 & 1.21, 1.22, 1.21 & 0.811, 0.841, 0.825 \\
NGC 6139 & 8965              & 17.7             & 2       & 1                & 0.260               \\
NGC 6144 & 7458              & 53.4             & 2       & 1                & 0.330               \\
NGC 6205 & 5436, 7290        & 26.5, 26.6       & 2, 2    & 1, 1             & 0.464               \\
NGC 6218 & 4530              & 22.8             & 2       & 1                & 0.447               \\
NGC 6287 & 13734             & 39.7             & 2       & 1                & 0.268               \\  
NGC 6304 & 11073, 8952       & 96.2, 4.9        & 2, 2    & 1.52, 1          & 0.143, 0.321        \\
NGC 6333 & 8954              & 7.8              & 2       & 1                & 0.288               \\
NGC 6341 & 5241, 3778        & 19.7, 29.3       & 2, 2    & 1, 1             & 0.246, 0.241        \\
NGC 6352 & 13674             & 19.5             & 2       & 1                & 0.595               \\
NGC 6362 & 11024, 12038      & 28.8, 8.7        & 1, 1    & 1, 1             & 0.411, 0.404        \\
NGC 6366 & 2678              & 20.1             & 1       & 1                & 0.784               \\
NGC 6388 & 5505              & 44.6             & 2       & 1                & 0.200               \\
NGC 6401 & 8948              & 11.1             & 2       & 1.04             & 0.578               \\
NGC 6402 & 8947              & 11.8             & 2       & 1                & 0.299               \\
NGC 6440 & 947, 3799, 11802  & 22.6, 23.4, 4.91 & 2, 2, 2 & 1, 1, 1          & 0.200, 0.200, 0.200 \\
NGC 6517 & 9597              & 23.3             & 2       & 1                & 0.200               \\
NGC 6528 & 8961, 12400       & 12.2, 52.7       & 2, 2    & 1, 1             & 0.200, 0.200        \\
NGC 6535 & 11025             & 52.4             & 2       & 1                & 0.200               \\
NGC 6539 & 8949              & 14.4             & 2       & 1                & 0.440               \\
NGC 6553 & 13671, 8957       & 30.9, 4.9        & 2, 2    & 1, 1             & 0.255, 0.285        \\
NGC 6569 & 8974              & 11.1             & 2       & 1                & 0.245               \\
NGC 6626 & 2683, 2684, 2685  & 8.9, 12.4, 12.5  & 2, 2, 2 & 1, 1, 1          & 0.632, 0.602, 0.606 \\
         & 9132, 9133        & 141, 52.5        & 2, 2    & 1, 1             & 0.568, 0.561        \\
NGC 6637 & 8946              & 7.4              & 2       & 1                & 0.253,              \\
NGC 6638 & 8950              & 8.3              & 2       & 1                & 0.200               \\
NGC 6656 & 5437, 14609       & 15.8, 84.6       & 1, 1    & 1.02, 1.26       & 1.219, 0.799        \\           
NGC 6715 & 4448              & 29.1             & 2       & 1.59             & 0.150               \\
NGC 6717 & 13733             & 16.9             & 2       & 1                & 0.240               \\
NGC 6760 & 13672             & 50.7             & 2       & 1                & 0.277               \\
NGC 6809 & 4531              & 33.7             & 1       & 1.04             & 0.622               \\
NGC 6838 & 5434              & 50.8             & 2       & 1                & 0.453               \\
NGC 7089 & 8960              & 11.2             & 2       & 1                & 0.326               \\
Glim 01  & 6587              & 44.6             & 2       & 1                & 0.200               \\
Pal 10   & 8945              & 10.8             & 2       & 1                & 0.299               \\
Ter 3    & 11026             & 64.3             & 2       & 1                & 0.231               \\
Ter 5    & 13706,10059,13705 & 45.5, 35.9, 13.2 & 2, 2, 2 & 1, 1, 1          & 0.225, 0.244, 0.243 \\
         & 14339,14625,15615 & 34.1, 48.9, 82.9 & 2, 2, 2 & 1, 1, 1          & 0.235, 0.262, 0.266 \\
         & 3798,13225,13252  & 24.9, 29.7, 39.2 & 2, 2 ,2 & 1, 1, 1          & 0.200, 0.234, 0.241 \\
         & 14475,14476,14477 & 29.3, 28.0, 28.3 & 2, 2, 2 & 1, 1, 1          & 0.426, 0.428, 0.428 \\
         & 14478             & 28.3             & 2       & 1                & 0.449               \\
\hline
\sidehead{\textbf{Core Collapse GCs:}}
NGC 362  & 4529, 5229        & 51.1, 8          & 2, 2    & 1, 1             & 0.218, 0.218        \\
NGC 1904 & 9027              & 10               & 2       & 1                & 0.200               \\
NGC 5946 & 9956              & 24.3             & 1       & 1                & 0.349               \\
NGC 6256 & 8951              & 9.4              & 2       & 1                & 0.268               \\
NGC 6266 & 2677              & 60.7             & 2       & 1                & 0.200               \\
NGC 6293 & 8962              & 9.6              & 2       & 1                & 0.271               \\
NGC 6325 & 8959              & 16.7             & 2       & 1                & 0.237               \\
NGC 6342 & 9957              & 15.8             & 2       & 1                & 0.236               \\
NGC 6355 & 9958              & 22.2             & 2       & 1                & 0.274               \\
NGC 6397 & 79, 2668, 2669    & 48.3, 28.1, 26.7 & 1, 1, 1 & 1.19, 1.04, 1.03 & 0.213, 0.708, 0.708 \\       
         & 7460, 7461        & 146, 88.9        & 1, 1    & 1.11, 1.10       & 0.596, 0.671        \\
NGC 6453 & 9959              & 20.4             & 2       & 1                & 0.200               \\
NGC 6522 & 8963              & 8.33             & 2       & 1                & 0.272               \\
NGC 6541 & 3779              & 43.2             & 2       & 1                & 0.274               \\
NGC 6544 & 5435              & 8.96             & 2       & 1                & 0.266               \\
NGC 6558 & 9961              & 11               & 1       & 1.08             & 0.516               \\
NGC 6642 & 8955              & 6.4              & 2       & 1                & 0.236               \\
NGC 6681 & 9955, 8958        & 67.8, 6.9        & 2, 2    & 1, 1             & 0.233, 0.237        \\
NGC 6752 & 948, 6612         & 27.5, 37         & 2, 2    & 1, 1             & 0.501, 0.585        \\
NGC 7099 & 2679              & 43               & 2       & 1                & 0.217               \\
Ter 1    & 5464              & 18.5             & 1       & 1.24             & 0.317               \\                                       
Ter 9    & 9960              & 15.2             & 2       & 1                & 0.239               \\
\enddata
\tablecomments{Column 1-3: Name, {\it Chandra} observation ID and effective exposure time. Column 4: background regions, 1 and 2 denote the use of annulus with radii $r_{h}-2r_{h}$ and $2r_{h}-3r_{h}$, respectively. Column 5-6: value of $S_{h}/S_{fov}$ and $S_{fov}/B_{fov}$.}
\label{tab:log}
\end{deluxetable}

\begin{deluxetable}{llllllll}
\tabletypesize{\scriptsize}
\tablecolumns{8}
\linespread{0.65}
\tablewidth{0pc}
\tablecaption{GCs with X-ray sources in outburst}
\tablehead{
\colhead{Name} & \colhead{ID} & \colhead{$N_{s}-N_b'$}  & \colhead{$S1$}  & \colhead{$L_{X,1}$} & \colhead{$L_{X}$} & \colhead{$L_{X,tot}$} & \colhead{$2L_{X,tot}/M$} \\
\colhead{(1)} & \colhead{(2)} & \colhead{(3)} & \colhead{(4)} & \colhead{(5)} & \colhead{(6)} & \colhead{(7)}  & \colhead{(8)} }
\startdata
NGC 5272 & 4542  & 419  & 297  & $3.02^{+0.23}_{-0.26}$ & $0.70^{+0.38}_{-0.26}$ & $3.63^{+0.41}_{-0.30}$ & $0.92^{+0.38}_{-0.33}$ \\
         & 4543  & 750  & 628  & $5.84^{+0.35}_{-0.62}$ & $1.05^{+0.29}_{-0.29}$ & $8.18^{+0.93}_{-0.80}$ & $2.07^{+0.86}_{-0.89}$ \\
NGC 6717 & 13733 & 865  & 643  & $5.49^{+0.23}_{-0.56}$ & $0.97^{+0.14}_{-0.15}$ & $5.82^{+0.35}_{-0.41}$ & $19.8^{+9.08}_{-9.28}$ \\
NGC 6453 & 9959  & 286  & 255  & $3.17^{+0.32}_{-0.28}$ & $0.46^{+0.13}_{-0.10}$ & $3.41^{+0.29}_{-0.37}$ & $4.37^{+0.37}_{-0.47}$ \\
\enddata
\vspace{-0.3cm}
\tablecomments{Column 1-2: Cluster name, {\it Chandra} observation ID. Column 3-4: Cluster net counts and counts of the outburst source. Column 5-8: Luminosities (0.5-8 keV) of the outburst source, GC cumulative luminosity when the outburst source has been rejected and included separatively, with a units of $10^{33}\rm\ erg\ s^{-1}$. Column 9: The total X-ray emissivity of GCs in units of $10^{28}\rm\ erg\ s^{-1}M_{\odot}^{-1}$.}
\label{tab:burst}
\end{deluxetable}

\clearpage
\begin{deluxetable}{lllllllllll}
\tabletypesize{\scriptsize}
\tablecolumns{11}
\tablewidth{0pc}
\tablecaption{DERIVED X-RAY PROPERTIES}
\tablehead{
\colhead{Name} & \colhead{Net counts} & \colhead{S/N} & \colhead{Model} & \colhead{nH} & \colhead{PI} & \colhead{KT} & \colhead{$\chi^{2}_{\nu}$} & \colhead{$L_{X,0.5-8}$} & \colhead{$L_{K}$} & \colhead{$L_{X}/L_{K}$} \\
\colhead{(1)} & \colhead{(2)} & \colhead{(3)} & \colhead{(4)} & \colhead{(5)} & \colhead{(6)} & \colhead{(7)} & \colhead{(8)} & \colhead{(9)} & \colhead{(10)} & \colhead{(11)}}
\startdata
\sidehead{\textbf{Normal GCs:}}
NGC 104$^*$ & 17246 & 102.5 &PL+Brem& 0.023 & $0.42^{+0.04}_{-0.05}$ &$0.65^{+0.05}_{-0.08}$& 376.3/352 & $58.9^{+7.65}_{-8.56}$ & 74.6  & $7.90^{+1.03}_{-1.15}$\\ 
NGC 2808    & 700   & 19.2  & PL    & 0.128 & $1.92^{+0.16}_{-0.16}$ &                      & 21.3/38   & $28.9^{+2.53}_{-1.86}$ & 46.91 & $6.16^{+0.54}_{-0.40}$\\
NGC 288$^*$ & 230   & 3.2   & PL    & 0.017 & $2.15^{+1.72}_{-1.51}$ &                      & 42.8/22   & $1.28^{+1.19}_{-1.16}$ & 3.42  & $3.74^{+3.48}_{-3.40}$\\ 
NGC 3201    & 93    & 0.48  & PL    & 0.139 & $2$(frozen)            &                      &           & $< 1.49$               & 7.21  & $< 2.07$ \\
NGC 5024    & 184   & 4.2   & PL    & 0.012 & $2$(frozen)            &                      & 3.8/3     & $12.1^{+2.56}_{-2.59}$ & 23.1  & $5.24^{+1.11}_{-1.12}$\\
NGC 5139$^*$& 12012 & 32.4  &PL+Brem& 0.069 & $1.52^{+0.16}_{-0.14}$ &$0.21^{+0.07}_{-0.08}$& 171.3/167 & $14.1^{+1.56}_{-1.09}$ & 112.2 & $1.26^{+0.14}_{-0.10}$\\
NGC 5272$^*$& 365   & 5.1   & PL    & 0.006 & $2.34^{+0.81}_{-0.64}$ &                      & 58.9/46   & $9.14^{+3.09}_{-2.63}$ & 32.98 & $2.77^{+0.94}_{-0.80}$\\ 
NGC 5286    & 134   & 6.1   & PL    & 0.139 & $2.1^{+0.52}_{-0.49}$  &                      & 2.8/8     & $13.5^{+3.55}_{-2.6}$  & 26.1  & $5.17^{+1.36}_{-0.99}$\\
NGC 5824    & 14    & 1.2   & PL    & 0.075 & $2.0$(frozen)          &                      &           & $< 27.8$               & 35.98 & $< 7.73$ \\
NGC 5904    & 513   & 6.9   & PL    & 0.017 & $1.8^{+0.42}_{-0.41}$  &                      & 9.5/10    & $5.06^{+1.37}_{-0.83}$ & 22.75 & $2.22^{+0.60}_{-0.37}$\\
NGC 5927    & 330   & 6.0   & PL    & 0.261 & $2.0^{+0.58}_{-0.47}$  &                      & 9.4/10    & $4.86^{+1.23}_{-0.83}$ & 17.25 & $2.82^{+0.71}_{-0.48}$\\
NGC 6093    & 960   & 23.3  & PL    & 0.104 & $1.78^{+0.12}_{-0.11}$ &                      & 44.4/52   & $23.7^{+1.39}_{-1.81}$ & 16.05 & $14.8^{+0.87}_{-1.13}$\\ 
NGC 6121$^*$& 3791  & 12.2  &PL+Brem& 0.203 & $0.60^{+0.43}_{-0.73}$ &$0.73^{+0.72}_{-0.30}$& 385.3/279 & $5.26^{+0.03}_{-0.70}$ & 6.14  & $8.56^{+0.05}_{-1.14}$\\
NGC 6139    & 288   & 9.1   & PL    & 0.435 & $2.30^{+0.37}_{-0.34}$ &                      & 9.3/13    & $24.9^{+3.88}_{-3.48}$ & 18.87 & $13.2^{+2.06}_{-1.84}$\\
NGC 6144    & 428   & 4.9   & PL    & 0.209 & $3.2$                  &                      & 11.7/6    & $4.95^{+1.35}_{-0.81}$ & 4.67  & $10.6^{+2.89}_{-1.73}$\\
NGC 6205    & 1373  & 15.0  & PL    & 0.012 & $2.14^{+0.17}_{-0.17}$ &                      & 50.4/51   & $9.15^{+0.62}_{-0.68}$ & 19.4  & $4.72^{+0.32}_{-0.35}$\\
NGC 6218    & 380   & 6.3   & PL    & 0.11  & $1.55^{+0.51}_{-0.49}$ &                      & 5.3/7     & $4.01^{+0.97}_{-0.81}$ & 5.1   & $7.87^{+1.90}_{-1.59}$\\
NGC 6287    & 232   & 6.9   & PL    & 0.348 & $1.15^{+0.41}_{-0.41}$ &                      & 5.4/9     & $9.75^{+1.61}_{-1.93}$ & 6.68  & $14.6^{+2.41}_{-2.89}$\\  
NGC 6304    & 1561  & 18.4  & PL    & 0.313 & $3.2^{+0.22}_{-0.21}$  &                      & 57.7/56   & $17.2^{+0.9}_{-0.6}$   & 11.55 & $14.9^{+0.78}_{-0.52}$\\
NGC 6333    & 23    & 1.1   & PL    & 0.22  & $2.0$(frozen)          &                      &           & $< 5.14$               & 10.58 & $< 4.86$ \\
NGC 6341    & 643   & 12.4  & PL    & 0.012 & $2.16^{+0.22}_{-0.22}$ &                      & 48.4/30   & $5.42^{+0.51}_{-0.46}$ & 13.23 & $4.10^{+0.38}_{-0.34}$\\
NGC 6352$^*$& 337   & 5.4   & PL    & 0.128 & $3.17^{+0.65}_{-0.69}$ &                      & 35.1/25   & $4.14^{+0.84}_{-0.87}$ & 6.07  & $6.82^{+1.38}_{-1.43}$\\
NGC 6362    & 106   & 1.1   & PL    & 0.052 & $2.0$(frozen)          &                      &           & $< 3.88$               & 6.14  & $< 6.32$ \\
NGC 6366$^*$& 445   & 4.9   & PL    & 0.412 & $1.93^{+0.64}_{-0.63}$ &                      & 68.6/57    &$3.17^{+1.04}_{-0.98}$ & 1.06  & $30.0^{+9.84}_{-9.28}$\\
NGC 6388    & 3682  & 54.9  & PL    & 0.215 & $2.05^{+0.05}_{-0.05}$ &                      & 176.2/136 & $88.8^{+1.67}_{-2.80}$ & 58.38 & $15.2^{+0.29}_{-0.48}$\\
NGC 6401    & 74    & 1.5   & PL    & 0.418 & $2.0$(frozen)          &                      &           & $< 23.2$               & 17.58 & $< 13.2$ \\
NGC 6402    & 125   & 3.7   & PL    & 0.348 & $0.99^{+0.9}_{-0.93}$  &                      & 1.44/3    & $17.7^{+6.38}_{-6.11}$ & 33.16 & $5.34^{+1.92}_{-1.84}$\\
NGC 6440    & 2672  & 46.9  & PL    & 0.621 & $2.78^{+0.08}_{-0.08}$ &                      & 133/122   & $72.5^{+1.73}_{-2.11}$ & 36.51 & $19.9^{+0.47}_{-0.58}$\\
NGC 6517    & 17    & 0.9   & PL    & 0.626 & $2.0$(frozen)          &                      &           & $< 4.53$               & 9.82  & $< 4.61$ \\
NGC 6528    & 262   & 8.7   & PL    & 0.313 & $1.98^{+0.33}_{-0.31}$ &                      & 7.1/14    & $3.73^{+0.59}_{-0.69}$ & 6.23  & $5.99^{+0.95}_{-1.11}$\\
NGC 6535    & 30    & 0.7   & PL    & 0.197 & $2.0$(frozen)          &                      &           & $< 1.15$               & 0.607 & $< 19.0$ \\
NGC 6539    & 333   & 6.8   & PL    & 0.592 & $3.37^{+0.69}_{-0.65}$ &                      & 10.3/11   & $31.7^{+6.41}_{-4.41}$ & 18.34 & $17.3^{+3.49}_{-2.40}$\\
NGC 6553    & 627   & 12.4  & PL    & 0.365 & $2.88^{+0.32}_{-0.31}$ &                      & 25.9/26   & $9.96^{+1.03}_{-0.63}$ & 32.1  & $3.10^{+0.32}_{-0.20}$\\
NGC 6569    & 18    & 0.8   & PL    & 0.307 & $2.0$(frozen)          &                      &           & $< 7.58$               & 18.17 & $< 4.17$\\ 
NGC 6626    & 37119 & 123.4 & PL    & 0.232 & $1.98^{+0.03}_{-0.03}$ &                      & 1135.5/658& $58.3^{+0.58}_{-0.53}$ & 19.75 & $29.5^{+0.29}_{-0.27}$\\
NGC 6637    & 99    & 4.8   & PL    & 0.104 & $1.0^{+0.53}_{-0.49}$  &                      & 4.7/5     & $16.6^{+5.57}_{-3.74}$ & 15.98 & $10.4^{+3.49}_{-2.34}$\\
NGC 6638    & 90    & 5.7   & PL    & 0.238 & $1.35^{+0.52}_{-0.48}$ &                      & 1.24/6    & $14.7^{+3.39}_{-2.44}$ & 7.51  & $19.6^{+4.51}_{-3.25}$\\
NGC 6656$^*$& 3629  & 17.6  & PL    & 0.197 & $1.66^{+0.69}_{-0.34}$ &                      & 223.5/155 & $4.70^{+2.18}_{-0.96}$ & 20.81 & $2.26^{+1.05}_{-0.46}$\\
NGC 6715    & 271   & 11.6  & PL    & 0.087 & $2.38^{+0.24}_{-0.23}$ &                      & 18.04/15  & $124^{+12.5}_{-7.08}$  & 117.3 & $10.6^{+1.07}_{-0.61}$\\
NGC 6717    & 221   & 8.9   & PL    & 0.128 & $0.99^{+0.33}_{-0.32}$ &                      & 1.49/10   & $9.67^{+1.48}_{-1.51}$ & 2.45  & $39.4^{+6.04}_{-6.16}$\\
NGC 6760    & 85    & 1.5   & PL    & 0.447 & $2.0$(frozen)          &                      &           & $< 3.03$               & 20.63 & $< 1.47$ \\
NGC 6809    & 10    & 0.09  & PL    & 0.046 & $2.0$(frozen)          &                      &           & $< 1.26$               & 7.02  & $< 1.80$ \\
NGC 6838    & 1063  & 12.9  & PL    & 0.145 & $1.97^{+0.29}_{-0.28}$ &                      & 15/18     & $2.67^{+0.51}_{-0.36}$ & 2.33  & $11.5^{+2.19}_{-1.55}$\\
NGC 7089    & 101   & 3.8   & PL    & 0.035 & $1.54^{+0.54}_{-0.48}$ &                      & 3.1/5     & $14.2^{+6.37}_{-4.01}$ & 33.16 & $4.28^{+1.92}_{-1.21}$\\
Glim 01     & 1107  & 25.3  & PL    & 2.813 & $1.58^{+0.21}_{-0.21}$ &                      & 65.9/48   & $15.6^{+0.72}_{-0.97}$ & 26.98 & $5.78^{+0.27}_{-0.36}$\\
Pal 10      & 15    & 0.7   & PL    & 0.963 & $2.0$(frozen)          &                      &           & $< 3.54$               & 2.98  & $< 11.9$ \\
Ter 3       & -106  & -1.5  & PL    & 0.423 & $2.0$(frozen)          &                      &           & $< 1.07$               & 0.37  & $< 29.0$ \\
Ter 5       & 39938 & 170.5 & PL    & 1.380 & $1.93^{+0.02}_{-0.02}$ &                      &2320.6/1687& $117^{+0.84}_{-1.13}$  & 96.65 & $12.1^{+0.09}_{-0.12}$\\
\hline
\sidehead{\textbf{Core Collapse GCs:}}
NGC 362     & 1368  & 26.3  & Brem  & 0.029 &                        &$1.15^{+0.16}_{-0.13}$& 62.7/61   & $9.20^{+0.34}_{-0.46}$ & 20.63 & $4.46^{+0.17}_{-0.22}$\\
NGC 1904    & 58    & 3.7   & PL    & 0.006 & $2.04^{+0.93}_{-0.72}$ &                      & 1.7/4     & $7.42^{+2.48}_{-2.01}$ & 7.83  & $9.48^{+3.17}_{-2.57}$\\
NGC 5946    & 16    & 0.5   & PL    & 0.313 & $2.0$(frozen)          &                      &           & $< 4.73$               & 15.94 & $< 2.97$ \\ 
NGC 6256    & 135   & 6.1   & PL    & 0.632 & $1.81^{+0.51}_{-0.46}$ &                      & 8.2/7     & $29.4^{+4.3}_{-8.09}$  & 11.64 & $25.3^{+3.69}_{-6.95}$\\
NGC 6266    & 5894  & 63.9  &PL+Brem& 0.273 & $1.15^{+0.29}_{-0.36}$ &$0.74^{+0.15}_{-0.13}$& 238.6/176 & $63.9^{+1.02}_{-4.99}$ & 39.86 & $16.0^{+0.27}_{-1.25}$\\
NGC 6293    & 49    & 2.1   & PL    & 0.209 & $2.0$(frozen)          &                      & 0/1       & $< 10.1$               & 10.67 & $< 9.47$\\
NGC 6325    & 82    & 4.1   & PL    & 0.528 & $2.0$(frozen)          &                      & 3.06/2    & $3.17^{+1.18}_{-0.90}$ & 3.79  & $8.36^{+3.11}_{-2.37}$\\
NGC 6342    & 84    & 3.5   & PL    & 0.267 & $0.97^{+0.66}_{-0.61}$ &                      & 1.5/4     & $8.47^{+2.97}_{-1.87}$ & 3.9   & $21.7^{+7.62}_{-4.80}$\\
NGC 6355    & 108   & 3.2   & PL    & 0.447 & $1.70^{+1.10}_{-0.84}$ &                      & 0.34/3    & $7.28^{+2.77}_{-2.56}$ & 9.47  & $7.69^{+2.92}_{-2.70}$\\
NGC 6397$^*$& 58210 & 133.5 &PL+Brem& 0.104 & $1.19^{+0.02}_{-0.02}$ &$0.29^{+0.08}_{-0.07}$& 1315/1274 & $13.3^{+0.48}_{-0.47}$ & 1.98  & $67.3^{+2.41}_{-2.38}$\\       
NGC 6453    & 284   & 12.2  & PL    & 0.371 & $2.0$(frozen)          &                      & 3.2/3     & $5.13^{+1.41}_{-1.14}$ & 7.2   & $7.13^{+1.96}_{-1.58}$\\
NGC 6522    & 64    & 2.5   & PL    & 0.278 & $2.0$(frozen)          &                      &           & $< 10.1$               & 11.48 & $< 8.8$\\
NGC 6541    & 2870  & 39.5  & PL    & 0.081 &                        &$1.43^{+0.13}_{-0.11}$& 139.3/103 & $22.6^{+0.07}_{-0.08}$ & 13.28 & $17.0^{+0.50}_{-0.59}$\\
NGC 6544    & 132   & 4.2   & PL    & 0.441 & $2.0$(frozen)          &                      & 3.67/3    & $1.71^{+0.44}_{-0.38}$ & 3.28  & $5.21^{+1.35}_{-1.15}$\\
NGC 6558    & 87    & 1.6   & PL    & 0.255 & $2.0$(frozen)          &                      &           & $< 12.62$               & 3.35  & $< 37.6$ \\
NGC 6642    & 19    & 1.2   & PL    & 0.232 & $2.0$(frozen)          &                      &           & $< 5.17$               & 4.06  & $< 12.7$ \\
NGC 6681    & 72    & 1.4   & PL    & 0.041 & $2.0$(frozen)          &                      &           & $< 1.38$               & 4.62  & $< 2.99$\\
NGC 6752    & 4640  & 35.4  & PL    & 0.023 & $1.47^{+0.06}_{-0.06}$ &                      & 221.7/153 & $11.1^{+0.38}_{-0.42}$ & 7.25  & $15.3^{+0.52}_{-0.58}$\\  
NGC 7099    & 1045  & 18.0  & PL    & 0.017 & $2.32^{+0.12}_{-0.12}$ &                      & 68.2/50   & $8.71^{+0.60}_{-0.56}$ & 5.22  & $16.7^{+1.15}_{-1.07}$\\
Ter 1       & 633   & 12.2  & PL    & 1.154 & $1.41^{+0.33}_{-0.32}$ &                      & 18.5/19   & $24.5^{+2.12}_{-3.06}$ & 27.34 & $8.96^{+0.77}_{-1.12}$\\
Ter 9       & 275   & 8.9   & PL    & 1.021 & $1.17^{+0.32}_{-0.34}$ &                      & 8.9/13    & $23.9^{+2.37}_{-3.06}$ & 6.26  & $38.2^{+3.79}_{-4.89}$\\
\enddata
\tablecomments{Column 1-3: Name, observation net counts (0.5-8 keV) and signal-to-noise ratio. Column 4-5: X-ray spectral fitting model and the fixed absorption column in units of $10^{22}\rm\ cm^{-2}$, which is calculated with $N_{H}=0.58 \times E(B-V) \times 10^{22}\rm\ {cm}^{-2}$. GCs marks with ``*" have a large angular extent ($r_h \gtrsim 2'$), we adopted the ``Double Subtraction" procedure to correct the vignetting effect during spectral fitting. Column 6-8: Photon-index, plasma temperature in units of keV, and $\chi^2$/degree of freedom. Column 9-11: 0.5-8 keV X-ray luminosity of GCs in units of $10^{32}\rm\ erg\ s^{-1}$, K-band luminosity in units of $10^{4}L_{K,\odot}$ and X-ray to K-band luminosity ratio, in units of $10^{27}\rm\ erg\ s^{-1}L_{K,\odot}^{-1}$.}
\label{tab:LX}
\end{deluxetable}

\begin{deluxetable}{lllllllll}
\tabletypesize{\scriptsize}
\tablecolumns{9}
\tablewidth{0pc}
\tablecaption{X-RAY VARIABILITY OF THREE GCs}
\tablehead{
\colhead{Obs.ID} & \colhead{Exp.} & \colhead{Date} & \colhead{Model} & \colhead{nH} & \colhead{Photon} & \colhead{} & \colhead{$\chi^{2}_{\nu}$}  & \colhead{$L_{X}$} \\
\colhead{} & \colhead{(ks)} & \colhead{} & \colhead{} & \colhead{($10^{22}cm^{-2}$)} & \colhead{Index} & \colhead{} & \colhead{(d.o.f.)} & \colhead{($10^{32}erg/s$)}}
\startdata
\sidehead{Terzan 5:}
3798   & 24.9 & 2003-07-13 & PL      & 1.322 & $1.82^{+0.08}_{-0.08}$ &                       & 116.7/109 & $137.3^{+3.85}_{-4.43}$  \\
10059  & 35.9 & 2009-07-15 & PL      & 1.322 & $1.81^{+0.10}_{-0.09}$ &                       & 129.3/135 & $100.4^{+2.50}_{-3.13}$  \\
13225  & 29.7 & 2011-02-17 & PL      & 1.322 & $1.99^{+0.09}_{-0.09}$ &                       & 101.9/116 & $119.2^{+3.62}_{-3.71}$  \\
13252  & 39.2 & 2011-04-29 & PL      & 1.322 & $1.98^{+0.09}_{-0.08}$ &                       & 151.2/122 & $101.0^{+3.28}_{-3.59}$  \\
13705  & 13.2 & 2011-09-05 & PL      & 1.322 & $1.99^{+0.14}_{-0.13}$ &                       & 66.8/57   & $116.0^{+4.67}_{-5.84}$  \\
13706  & 45.5 & 2012-05-13 & PL      & 1.322 & $1.82^{+0.06}_{-0.06}$ &                       & 202.9/187 & $145.6^{+2.82}_{-3.17}$  \\                    
14339  & 31.4 & 2011-09-08 & PL      & 1.322 & $1.98^{+0.09}_{-0.09}$ &                       & 120.1/121 & $111.7^{+2.61}_{-3.89}$  \\
14475  & 29.3 & 2012-09-17 & PL      & 1.322 & $2.15^{+0.10}_{-0.10}$ &                       & 111.2/105 & $116.4^{+3.42}_{-3.54}$  \\
14476  & 28.0 & 2012-10-28 & PL      & 1.322 & $2.03^{+0.10}_{-0.10}$ &                       & 112.7/99  & $110.2^{+4.01}_{-3.39}$  \\
14477  & 28.3 & 2013-02-05 & PL      & 1.322 & $2.02^{+0.10}_{-0.10}$ &                       & 117.9/100 & $112.5^{+3.08}_{-5.12}$  \\
14478  & 28.3 & 2013-07-16 & PL      & 1.322 & $1.88^{+0.09}_{-0.09}$ &                       & 103.7/104 & $115.4^{+3.83}_{-3.76}$  \\
14625  & 48.9 & 2013-02-22 & PL      & 1.322 & $1.91^{+0.07}_{-0.07}$ &                       & 229/175   & $125.1^{+2.40}_{-2.49}$  \\
15615  & 82.9 & 2013-02-23 & PL      & 1.322 & $1.88^{+0.05}_{-0.05}$ &                       & 336.4/236 & $113.1^{+1.72}_{-1.59}$  \\
Joint  & -    & --         & PL      & 0.232 & $1.93^{+0.02}_{-0.02}$ &                       &2320.6/1687& $116.9^{+0.84}_{-1.13}$  \\
\cline{1-9}
\sidehead{NGC 6626:}
2683  & 8.9  & 2002-09-09 & PL      & 0.232 & $2.03^{+0.15}_{-0.15}$ &                       & 52.8/52   & $45.3^{+2.11}_{-0.28}$  \\                    
2684  & 12.4 & 2002-07-04 & PL      & 0.232 & $1.93^{+0.12}_{-0.12}$ &                       & 59.3/65   & $45.0^{+2.59}_{-1.61}$  \\
2685  & 12.5 & 2002-08-04 & PL      & 0.232 & $2.14^{+0.14}_{-0.13}$ &                       & 50.3/62   & $38.4^{+2.28}_{-2.04}$  \\
9132  & 141  & 2008-08-07 & PL      & 0.232 & $1.96^{+0.03}_{-0.03}$ &                       & 339.5/296 & $63.8^{+0.90}_{-0.82}$  \\
9133  & 52.5 & 2008-08-10 & PL      & 0.232 & $2.00^{+0.06}_{-0.06}$ &                       & 261/185   & $58.8^{+1.17}_{-1.24}$  \\
Joint & --   & --         & PL      & 0.232 & $1.98^{+0.03}_{-0.03}$ &                       &1135.5/658 & $58.3^{+0.58}_{-0.53}$  \\
\cline{1-9} 
\sidehead{NGC 6397:}
79    & 48.3 & 2000-07-31 & PL      & 0.104 & $1.42^{+0.04}_{-0.04}$ &                       & 226.8/234 & $16.0^{+0.28}_{-0.33}$  \\
2668  & 28.1 & 2002-05-13 & PL      & 0.104 & $1.19^{+0.06}_{-0.06}$ &                       & 202.4/202 & $13.8^{+0.27}_{-0.37}$  \\
2669  & 26.7 & 2002-05-15 & PL      & 0.104 & $1.21^{+0.07}_{-0.07}$ &                       & 227.2/184 & $11.4^{+0.37}_{-0.27}$  \\
7460  & 146  & 2007-07-16 & PL      & 0.104 & $1.20^{+0.03}_{-0.03}$ &                       & 302.3/335 & $13.0^{+0.15}_{-0.16}$  \\
7461  & 88.9 & 2007-06-22 & PL      & 0.104 & $1.14^{+0.03}_{-0.03}$ &                       & 349.9/319 & $15.9^{+0.31}_{-0.18}$  \\
\enddata
\vspace{-0.5cm}
\tablecomments{Column 1-3: {\it Chandra} observation ID, effective exposure time in units of keV and observation date. Column 4-6: X-ray spectral fitting model, the fixed absorption column in units of $10^{22}\rm\ cm^{-2}$ and Photon-index. Column 8-9: $\chi^2$/degree of freedom and GC 0.5-8 keV X-ray luminosity in units of $10^{32}\rm\ erg\ s^{-1}$.}
\label{tab:var}
\end{deluxetable}

\begin{deluxetable}{lrrrrrrrrr}
\tabletypesize{\scriptsize}
\tablecolumns{10}
\tablewidth{0pc}
\tablecaption{TESTED CORRELATIONS}
\tablehead{
\multicolumn{1}{c}{} & \colhead{} & \multicolumn{2}{c}{Total GCs} &\colhead{} & \multicolumn{2}{c}{Normal GCs} & \colhead{} & \multicolumn{2}{c}{Core-Collapsed GCs}\\
\cline{3-4} \cline{6-7} \cline{9-10} \\
\colhead{Relations} & \colhead{} & \colhead{$r$} & \colhead{$p(>|r|)$} & \colhead{} & \colhead{$r$} & \colhead{$p(>|r|)$} & \colhead{} & \colhead{$r$} & \colhead{$p(>|r|)$} }
\startdata
$L_{X}-L_{K}$        & & 0.694    & $7.2\times10^{-11}$ & & 0.754    & $5.9\times10^{-10}$    & & 0.366      & 0.123 \\
$L_{X}-\Gamma$       & & 0.627    & $1.4\times10^{-8 }$ & & 0.769    & $1.7\times10^{-10}$    & & 0.194      & 0.426  \\
$\Gamma-M$           & & 0.611    & $4.1\times10^{-8 }$ & & 0.754    & $5.4\times10^{-11}$    & & 0.660      & 0.002  \\
$\gamma-M$           & & 0.111    & 0.374               & & 0.337    & 0.019                  & & 0.125      & 0.611  \\
$L_{X}/L_{K}-L_{K}$  & & -0.321   & 0.008               & & -0.272   & 0.061                  & & -0.228     & 0.348  \\
$L_{X}/L_{K}-\gamma$ & & 0.300    & 0.013               & & 0.298    & 0.040                  & & -0.149     & 0.542  \\
$L_{X}-Fe/H$         & & 0.043    & 0.734               & & 0.061    & 0.685                  & & 0.015      & 0.952  \\
$L_{X}/L_{K}-Fe/H$   & & 0.132    & 0.291               & & 0.234    & 0.114                  & & 0.014      & 0.955  \\
$L_{X}-\sigma$       & & 0.781    & $2.8\times10^{-9 }$ & & 0.867    & $2.0\times10^{-8 }$    & & 0.370      & 0.174  \\
$L_{X}/L_{K}-\sigma$ & & -0.055   & 0.737               & & -0.042   & 0.844                  & & -0.089     & 0.751  \\
$L_{X}-\rho$         & & 0.422    & $3.8\times10^{-4 }$ & & 0.721    & $7.2\times10^{-9 }$    & & 0.008      & 0.974  \\
$L_{X}/L_{K}-\rho$   & & 0.291    & 0.017               & & 0.234    & 0.109                  & & 0.053      & 0.830  \\
$L_{X}-c$            & & 0.317    & 0.009               & & 0.584    & $1.3\times10^{-5 }$    & & -0.019     & 0.938  \\
$L_{X}/L_{K}-c$      & & 0.247    & 0.044               & & 0.138    & 0.358                  & & 0.123      & 0.615  \\
$L_{X}-t_{c}$        & & -0.124   & 0.320               & & -0.331   & 0.023                  & & 0.184      & 0.450  \\
$L_{X}/L_{K}-t_{c}$  & & -0.423   & $4.0\times10^{-4}$  & & -0.457   & 0.001                  & & -0.191     & 0.432  \\
$L_{X}-t_{h}$        & & 0.270    & 0.027               & & 0.238    & 0.103                  & & 0.542      & 0.016  \\
$L_{X}/L_{K}-t_{h}$  & & -0.390   & $1.1\times10^{-3}$  & & -0.456   & 0.001                  & & 0.191      & 0.435  \\
$L_{X}-t$            & & 0.062    & 0.696               & & 0.019    & 0.921                  & & 0.275      & 0.414  \\
$L_{X}/L_{K}-t$      & & 0.376    & 0.014               & & 0.457    & 0.010                  & & 0.545      & 0.083  \\
$L_{X}-t_{r}$        & & -0.021   & 0.906               & & 0.014    & 0.942                  & & -0.152     & 0.774  \\
$L_{X}/L_{K}-t_{r}$  & & 0.417    & 0.013               & & 0.474    & 0.009                  & & 0.152      & 0.774  \\
$L_{X}-r_{gc}$       & & -0.093   & 0.454               & & -0.117   & 0.426                  & & -0.066     & 0.789  \\
$L_{X}/L_{K}-r_{gc}$ & & -0.402   & $7.5\times10^{-4}$  & & -0.454   & 0.001                  & & -0.134     & 0.587  \\
$L_{X}-e$            & & -0.118   & 0.374               & & -0.058   & 0.713                  & & -0.233     & 0.386  \\
$L_{X}/L_{K}-e$      & & -0.161   & 0.222               & & -0.210   & 0.176                  & & 0.009      & 0.974  \\
$L_{X}-r_t$          & & 0.103    & 0.405               & & 0.166    & 0.260                  & & 0.159      & 0.516  \\
$L_{X}/L_{K}-r_t$    & & -0.321   & 0.008               & & -0.398   & 0.005                  & & -0.307     & 0.201  \\
$L_{X}-f_{b}$        & & -0.721   & $6.9\times10^{-6 }$ & & -0.742   & $3.3\times10^{-5}$     & & -0.580     & 0.228  \\
$L_{X}/L_{K}-f_{b}$  & & -0.097   & 0.609               & & 0.114    & 0.597                  & & -0.493     & 0.321  \\
\enddata
\vspace{-0.5cm}
\tablecomments{Spearman's rank correlation test for each parameter. The $p(>|r|)$ values show the probability that a correlation arises randomly.}
\label{tab:spearman}
\end{deluxetable}

\newpage
\clearpage
\label{lastpage}


\begin{thebibliography}{}
\bibitem[Albrow et al.(2001)]{albrow2001} Albrow, M. D., Gilliland, R. L., Brown, T. M., Edmonds, P. D., et al. 2001, \apj, 559, 1060
\bibitem[Bacon et al.(1996)]{bacon1996} Bacon, D., Sigurdsson, S., Davis, M. B., 1996, \mnras, 281, 830
\bibitem[Bahramian et al.(2013)]{bahramian2013} Bahramian, A., Heinke, C.O., Sivakoff, G.R., \& Gladstone, J.C. 2013, \apj, 766, 136
\bibitem[Bellazzini et al.(1995)]{bellazzini1995} Bellazzini, M., Pasquali, A., Federici, L., et al. 1995, \apj, 439, 687
\bibitem[Cackett et al.(2006)]{cackett2006} Cackett, E. M., Wijnands, R., Heinke, C. O., Pooley, D., et al. 2006, \mnras, 369, 407
\bibitem[Chatterjee et al.(2010)]{chatterjee2010} Chatterjee, S., Fregeau, J. M., Umbreit, S., Rasio, F. A. 2010, \apj, 719, 915
\bibitem[{Clark}(1975)]{Clark1975} Clark, G. W. 1975, \apj, 199, L143
\bibitem[Cote et al.(1996)]{Cote1996} Cote, P., Fischer, P., 1996, \aj, 112,565
\bibitem[Davis et al.(2008)]{davis2008} Davis, D. S., Richer, H. B., Anderson, J., Brewer, J., et al. 2008, \aj, 135, 2141
\bibitem[Djorgovski \& Meylan(1994)]{djorgovski1994} Djorgovski, S., Meylan, G. 1994, \apj, 108, 1292
\bibitem[Edmonds et al.(2003)] {edmonds2003} Edmonds, P. D., Gilliland, R. L., Heinke, C. O., \& Grindlay, J. E. 2003, \apj, 569, 1177
\bibitem[Fabian et al.(1975)] {fabian1975} Fabian, A. C., Pringle, J. E., \& Rees, M. J. 1975, \mnras, 172, 15
\bibitem[Fischer \& Marcy(1992)]{fischer1992} Fischer, D.~A., \& Marcy, G.~W.\ 1992, \apj, 396, 178
\bibitem[Forbes \& Bridges(2010)]{forbes2010} Forbes, D. A., Bridges, T., 2010, \mnras, 404,1203
\bibitem[Fregeau et al.(2003)] {fregeau2003} Fregeau, J. M. et al. 2003, \apj, 593, 772
\bibitem[Fregeau(2008)] {fregeau2008} Fregeau, J. M. 2008, \apj, 673, 25
\bibitem[Fregeau(2009)] {fregeau2009} Fregeau, J. M., Ivanova, N., Rasio, F. A. 2009, \apj, 707, 1533
\bibitem[Gao et al.(1991)] {gao1991} Gao, B., Goodman, J., Cohn, H., Murphy, B. 1991, \apj, 370, 567
\bibitem[Ge et al.(2015)] {ge2015} Ge, C., Li, Z-Y., Xue, X-J., Gu, Q-S., et al. 2015, \apj, 812, 130
\bibitem[Gnedin et al.(2002)]{gnedin2002} Gnedin, O. Y., Zhao, H-S., et al. 2002, \mnras, 568, 23
\bibitem[Goodman \& Hut (1989)] {goodman1989} Goodman, J., Hut, P. 1989, \nat, 339, 40
\bibitem[Grindlay et al. (2001)] {grindlay2001} Grindlay, J. E., Heinke, C. O., Edmonds, P. D., \& Murray, S. S. 2001, science, 292, 2290
\bibitem[Haggard et al. (2009)] {haggard2009} Haggard, D., Cool, A. M., \& Davies, M. B. 2009, \apj, 697, 224
\bibitem[Harris(2010 edition)] {harris1996} Harris, W. E. 1996(2010 edition), \aj, 112, 1487.
\bibitem[Heggie(1975)]{heggie1975} Heggie, D. C., 1975, \mnras, 173, 729
\bibitem[Heinke(2011)]{heinke2011} Heinke, C. O. 2011, arXiv:astro-ph/1101.5356
\bibitem[Heinke et al.(2003)]{heinke2003} Heinke, C. O., Grindlay, J. E., Lugger, P. M., et al. 2003, \apj, 598, 501
\bibitem[Heinke et al.(2005)]{heinke2005} Heinke, C. O., Grindlay, J. E., Cohn, H. N., et al. 2005, \apj, 625, 796
\bibitem[Heinke(2006)] {heinke2006} Heinke, C.O., Wijnands, R., Cohn,H.N., et al. 2006, \apj, 651, 1098
\bibitem[Hills(1975)] {hills1975} Hills, J. G., 1975, \aj, 80, 809
\bibitem[Hills(1976)] {hills1976} Hills, J. G., 1976, \mnras, 175, 1P
\bibitem[Hut et al.(1992a)]{hut1992a} Hut, P., McMillan, S., Romani, R. W., 1992, \apj, 389, 527
\bibitem[Hut et al.(1992b)]{hut1992b} Hut, P., McMillan, S., Goodman, J., et al. 1992, \pasp, 104, 981
\bibitem[Hut (1993)]{hut1993} Hut, P., 1993, \apj, 403, 256
\bibitem[Ivanova et al.(2005)] {ivanova2005} Ivanova, N., Belczynski, K. Fregeau, J. M. and Rasio, F. A. 2005, \mnras, 358, 572
\bibitem[Ivanova(2006)] {ivanova2006} Ivanova, N. 2006, \apj, 636, 979
\bibitem[{Ivanova}(2011)]{ivanova2011} Ivanova, N. 2011, arXiv:astro-ph/1101.2684
\bibitem[Jarrett et al.(2000)]{jarrett2000} Jarrett, T. H., Chester, T., Cutri, R., et al. 2000, \aj, 119, 2498 
\bibitem[Ji \& Bregman(2015)]{ji2015} Ji, J., Bregman, J. N., 2015, \apj, 807, 32
\bibitem[Katz(1975)]{Katz1975} Katz, J. I. 1975, \nat, 253, 698
\bibitem[Kim et al.(2006)]{kim2006} Kim, E., Kim, D.-W., Fabbiano, G., et al. 2006, ApJ, 647, 276
\bibitem[Kim et al.(2013)] {kim2013} Kim, D. W., Fabbiano, N., Ivanova, Fragos, T., et al. 2013, \apj, 764, 98
\bibitem[King(1962)] {king1962} King, I. R. 1962, \aj, 67, 471
\bibitem[King(1966)] {1966} King, I. R. 1966, \aj, 71, 64
\bibitem[Kundu et al.(2002)] {kundu2002} Kundu, A., Maccarone, T. J., \& Zepf, S. E. 2002, \apjl, 574, 5
\bibitem[Lanzoni et al. (2010)] {lanzoni2010} Lanzoni, B., Ferraro, F. R., Dalessandro, E., et al. 2010, \apj, 717, 653
\bibitem[Leigh et al.(2015)]{leigh2015} Leigh, N. W. C., Giersz, M., et al. 2015, \mnras, 446, 226
\bibitem[Li et al.(2010)]{li2010} Li, Z., Spitler, L. R., Jones, C., et al. 2010, \apj, 721, 1368
\bibitem[Li et al.(2011)]{li2011} Li, Z., Jones, C., Forman,  W. R., et al. 2011, \apj, 730, 84
\bibitem[Liu et al.(2007)]{liu2007} Liu, Q. Z., van Paradijs, J., \& van den Heuvel, E. P. J., 2007, \aa, 469, 807
\bibitem[Lugger et al.(2007)] {lugger2007} Lugger, P. M., Cohn, H. N., Heinke, C. O., et al. 2007, \apj, 657, 286
\bibitem[Maccarone et al.(2004)] {maccarone2004} Maccarone, T. J., Kundu, A., Zepf, S. E. 2004, \apj, 606, 430
\bibitem[Ma et al.(2015)] {majun2015} Ma, J., Wang, S., Wu, Z-Y., et al. 2015, \aj, 149, 56
\bibitem[Marín-Franch et al.(2009)]{Marín-Franch2009} Marín-Franch, A., Aparicio, A., Piotto, G., et al. 2009, \apj, 694, 1498
\bibitem[Maxwell et al. (2012)] {maxwell2012} Maxwell, J. E., Lugger, P. M., Cohn, H. N., et al. 2012, \apj, 756, 147 
\bibitem[Mikkola (1983)]{mikkola1983} Mikkola, S., 1983, \mnras, 203, 1107
\bibitem[Milone et al.(2012)] {milone2012} Milone, A. P., Piotto, G., Bedin, L. R., et al. 2012, \aap, 540, 16
\bibitem[Paolillo et al.(2011)] {paolillo2011} Paolillo, M., Puzia, T. H., Goudfrooij, P., et al. 2011, \apj, 736, 90
\bibitem[Pooley et al.(2002a)] {pooley2002a} Pooley, D., Lewin, W. H. G., Homer L., et al. 2002, \apj, 569, 405
\bibitem[{{Pooley et al.}(2002b)}] {pooley2002b} Pooley, D., Lewin, W. H. G., Verbunt F., et al. 2002, \apj, 573, 184
\bibitem[{{Pooley et al.}(2003)}] {pooley2003} Pooley, D., et al. 2003, \apj, 591, 131
\bibitem[{Townsley \& Bildsten}(2005)]{Townsley2005} Townsley, D. M., Bildsten, L. 2005, \apj, 628,395
\bibitem[{{Pooley \& Hut}(2006)}] {pooley2006} Pooley, D., \& Hut, P. 2006, \apj, 646, 143
\bibitem[{{Pooley et al.}(2007)}] {pooley2007} Pooley, D., Rappaport, S., Levine, A., et al. 2007, arXiv: astro-ph/0708.3365
\bibitem[{{Pooley}(2010)}] {pooley2010} Pooley, D. 2010, PNAS, 107, 7164
\bibitem[{{Revnivtsev et al.}(2007)}] {Revnivtsev2007} Revnivtsev, M., Churazov, E., Sazonov, S., Forman, W., \& Jones,C. 2007, \aap, 473, 783
\bibitem[{{Sarazini et al.}(2003)}] {sarazin2003} Sarazin, C. L., Kundu, A., Irwin, J. A., et al. 2003, \apj, 595, 743
\bibitem[{{Sazonov et al.}(2006)}] {Sazonov2006} Sazonov, S., Revnivtsev, M., Gilfanov, M., et al. 2006, \aap, 450, 117
\bibitem[{{Sivakoff et al.}(2007)}] {sivakoff2007} Sivakoff, G. R., Jordan, A., Sarazin, C. L., et al. 2007, \apj, 660, 1246
\bibitem[{{Sollima et al.}(2007)}] {sollima2007} Sollima, A., Beccari, G., Ferraro, F. R., et al. 2007, \mnras, 380, 781
\bibitem[{{Sollima et al.}(2010)}] {sollima2010} Sollima, A., Carballo-Bello, J. A., Beccari, G., et al. 2010, \mnras, 401, 577
\bibitem[{{Sutantyo}(1975)}] {sutantyo1975} Sutantyo, W. 1975, \aap, 44, 227
\bibitem[{{Townsley \& Bildsten}(2005)}] {townsley2005} Townsley, D. M., \& Bildsten, L. 2005, \apj, 628, 395
\bibitem[{{van den Berg et al.}(2013)}] {vandenberg2013} van den Berg, M., Verbunt, F., Tagliaferri, G., et al. 2013, \apj, 770, 98
\bibitem[{{Verbunt \& Hut}(1987)}] {verbunt1987} Verbunt, F., \& Hut, P. 1987, in: The Origin and Evolution of Neutron Stars IAU Symp.125, eds. D.J. Helfand and J.H. Huang, Reidel, p.187
\bibitem[Verbunt(2000)]{verbunt2000} Verbunt, F. 2000, vol.198 of ASP Conf.Ser.,p.421
\bibitem[Verbunt(2001)]{verbunt2001} Verbunt, F. 2001, \aap, 368, 137
\bibitem[Verbunt(2003)]{verbunt2003} Verbunt, F. 2003, New Horizons in Globular Cluster Astronomy, ASP Conference Proceedings, Vol. 296, p. 245
\bibitem[Verbunt \& Freire (2014)]{verbunt2014} Verbunt, F., Freire, P. C. C. 2014, \aap, 561, 11
\end{thebibliography}
\end{document}